\documentclass{article}
\usepackage{amsthm}
\usepackage{graphicx}
\newcommand\maath{\mathsurround=0pt}
\newcommand{\EQM}[1]{\vcenter{\normalbaselines\maath
    \ialign{${\displaystyle ##}$\hfil&&\ ${\displaystyle ##}$\hfil\crcr
    \mathstrut\crcr\noalign{\kern-\baselineskip}
    \noalign{\smallskip}
    #1\crcr\mathstrut\crcr\noalign{\kern-\baselineskip}}}}
\newdimen\hoogte    \hoogte=12pt    
\newdimen\breedte   \breedte=14pt   
\newdimen\dikte     \dikte=0.5pt    
\def\beginYoung{
       \begingroup
       \def\vr{\vrule height0.8\hoogte width\dikte depth 0.3\hoogte}
       \def\fbox##1{\vbox{\offinterlineskip
                    \hrule height\dikte
                    \hbox to \breedte{\vr\hfill##1\hfill\vr}
                    \hrule height\dikte}}
       \vbox\bgroup \offinterlineskip \tabskip=-\dikte \lineskip=-\dikte
            \halign\bgroup &\fbox{##\unskip}\unskip  \crcr }
\def\End@Young{\egroup\egroup\endgroup}
\newenvironment{Young}{\beginYoung}{\End@Young}

\newdimen\breedtelong   \breedtelong=28pt   

\def\beginYounglong{
       \begingroup
       \def\vr{\vrule height0.8\hoogte width\dikte depth 0.3\hoogte}
       \def\fboxlong##1{\vbox{\offinterlineskip
                    \hrule height\dikte
                    \hbox to \breedtelong{\vr\hfill##1\hfill\vr}
                    \hrule height\dikte}}
       \vbox\bgroup \offinterlineskip \tabskip=-\dikte \lineskip=-\dikte
            \halign\bgroup &\fboxlong{##\unskip}\unskip  \crcr }
\def\Endlong@Younglong{\egroup\egroup\endgroup}
\newenvironment{Younglong}{\beginYounglong}{\Endlong@Younglong}

\def \1{{\rm Id}}
\def\grad{{\rm grad}}
\def\N{{{\rm I}\!{\rm N}}}
\def\R{{{\rm I}\!{\rm R}}}
\def\C{{\,\,\vrule depth0pt \!\!{\rm C}}}
\def\Q{{\,\,\vrule depth0pt \!\!{\rm Q}}}
\def\lk{R}
\def\lka{{\cal R}}
\def\la{\langle}
\def\ra{\rangle}
\def\lint{{\scriptstyle{\rfloor}}}

\begin{document}

\centerline{\bf Projective dynamics and first integrals}
\centerline {Alain Albouy, Alain.Albouy@obspm.fr}
\centerline {IMCCE-CNRS-UMR}
\centerline{Observatoire de Paris}
\centerline {77, avenue Denfert-Rochereau, 75014 Paris}
\centerline {France}
\bigskip\bigskip

{\bf Abstract.} We present the theory of tensors with Young tableau symmetry as an efficient computational tool in dealing with the polynomial first integrals of a natural system in classical mechanics. We relate a special kind of such first integrals, already studied by Lundmark, to Beltrami's theorem about projectively flat Riemannian manifolds. We set the ground for a new and simple theory of the integrable systems having only quadratic first integrals. This theory begins with two centered quadrics related by central projection, each quadric being a model of a space of constant curvature. Finally, we present an extension of these models to the case of degenerate quadratic forms.

\bigskip
\centerline{\bf 1. Introduction}
\bigskip

The dynamical systems defining the motion of a point $q$ in an affine space $A$ under a force field $f$ are of primary importance. They include the free fall of a particle, as considered by Galileo, and the most fundamental systems presented in Newton's book, the {\it Principia}. They are modeled by a connected open set ${\cal U}\subset A$, a vector field $f$ on ${\cal U}$, and the differential equation $${d^2q\over dt^2}=f,\eqno(1.1)$$
which we also write $\ddot q=f$. Such a system is defined without endowing $A$ with a Euclidean structure. But suppose there is a nondegenerate scalar product\footnote{What we call a {\it scalar product} is a field of symmetric bilinear forms on the tangent space, possibly degenerate.} $g$ on ${\cal U}$ such that

(i) $g$ is invariant by translation and

(ii) there is a function $U:{\cal U}\to\R$ such that $f=\grad_g U$.

\noindent Then, as is well-known, $L=T+U$, where $2T=g(\dot q,\dot q)$, is a Lagrangian for the system, and the energy $E=T-U$ is a conserved quantity (also called a first integral). Furthermore, there is a symplectic form defined in the phase space and invariant by the flow of the system.

Such a rich structure may also appear in the same elementary way after considering a change of the time parameter $t$. Define a new time $s$, through a positive function $\mu:{\cal U}\to\R$ and the formula:
$${d\over dt}=\mu{d\over ds}.\eqno(1.2)$$
We will often denote the new time derivative by $'$, and write for example $\dot q=\mu q'$. After this change of time and a division by $\mu^2$, $(1.1)$ becomes
$${1\over \mu}{d\over ds}\bigl(\mu{dq\over ds}\bigr)={f\over \mu^2}.\eqno(1.3)$$
After this division by $\mu^2$ the left-hand side is $q''+\mu^{-1}\sum_j(\partial \mu/\partial q_j)q'_jq'$. Due to the unit coefficient in front of $q''$, this expression is of the form $D_{q'}q'$, where $D$ is a symmetric (also called torsion-free) linear connection on the tangent bundle of ${\cal U}\subset A$. In usual index notation, where one writes $q^i$ instead of $q_i$, the Christoffel symbols of this connection are $\Gamma_{ij}^k=\omega_i\delta_j^k+\omega_j\delta_i^k$, where $2 \omega_i=\mu^{-1}(\partial\mu/\partial q^i)$, and where $\delta$ is the Kronecker $\delta$. Now, if there is  a nondegenerate scalar product $g$ on ${\cal U}$ such that

(i) $Dg=0$ and

(ii) there is a function $U:{\cal U}\to\R$ such that $\mu^{-2}f=\grad_g U$

\noindent then we get the same conclusions as in the previous case: $L=T+U$, where $2T=g(q',q')$, is a Lagrangian, $E=T-U$ is conserved and a symplectic form is conserved.

These considerations raise the following question: given a domain ${\cal U}$ and a force field $f$, which are the $\mu$'s such that there is a $g$ with properties (i) and (ii)? If we find two distinct pairs $(\mu, g)$, the system has two energies, which means in particular two quadratic first integrals (where quadratic means of degree two in the velocities). Lundmark's claim \cite{Lu1} {\it Two quadratic first integrals imply integrability} is relevant in this situation. Let us recall that this claim is not limited to systems with two degrees of freedom. {\it Newton systems} $(1.1)$ with two quadratic first integrals are concrete examples of {\it quasi-bi-Hamiltonian systems} in the sense of \cite{BCR}, the prefix {\it quasi} referring to the change of time (see also \cite{Ped}, \cite{CS2}). This often implies integrability {\it \`a la Liouville}.

Some of the possible $(\mu, g)$  appear special compared to the general ones: those with a constant $\mu$. But indeed the other possibilities are not so different. The geodesics of the connection $D$ describe straight lines, on which the length of the velocity vector may vary. In other words, $D$ is {\it geodesically equivalent} to the affine connection of the affine space $A$. The space $({\cal U}, D)$ is {\it projectively flat}. Now, we assume $Dg=0$, which means that $D$ is the Levi-Civita connection of $g$. A theorem by Beltrami applies: The open domain $({\cal U}, g)$ has constant curvature. Finally, a solution $(\mu,g, U)$ of our problem always corresponds to a natural system on a space of constant curvature.

Beltrami's theorem does not only constrain the intrinsic geometry of a projectively flat space. It also describes any admissible scalar product $g$ on ${\cal U}$ as a pull-back under some map. This map is a central projection, in a finite dimensional real vector space, from the affine hyperplane $A$ to a centered quadric. This quadric is a model for a pseudo-Riemannian space of constant curvature.

We get a model of an integrable system: two natural Lagrangian systems on two centered quadrics, each endowed with the standard pseudo-Riemannian scalar product of constant curvature, and corresponding to each other through central projection.  The zero curvature case is included: the quadric is then a pair of diametrically opposite hyperplanes, and the scalar product is invariant by translation. The first examples of such a model were described by Appell in 1890 (see \cite{Ap1}, \cite{Ap2}, \cite{Alb}, \cite{BorM}). His most striking example consists of the pair formed by the Kepler problem on a plane and the Kepler-Serret problem on a sphere.

\medskip
\centerline{\includegraphics [width=100mm]
{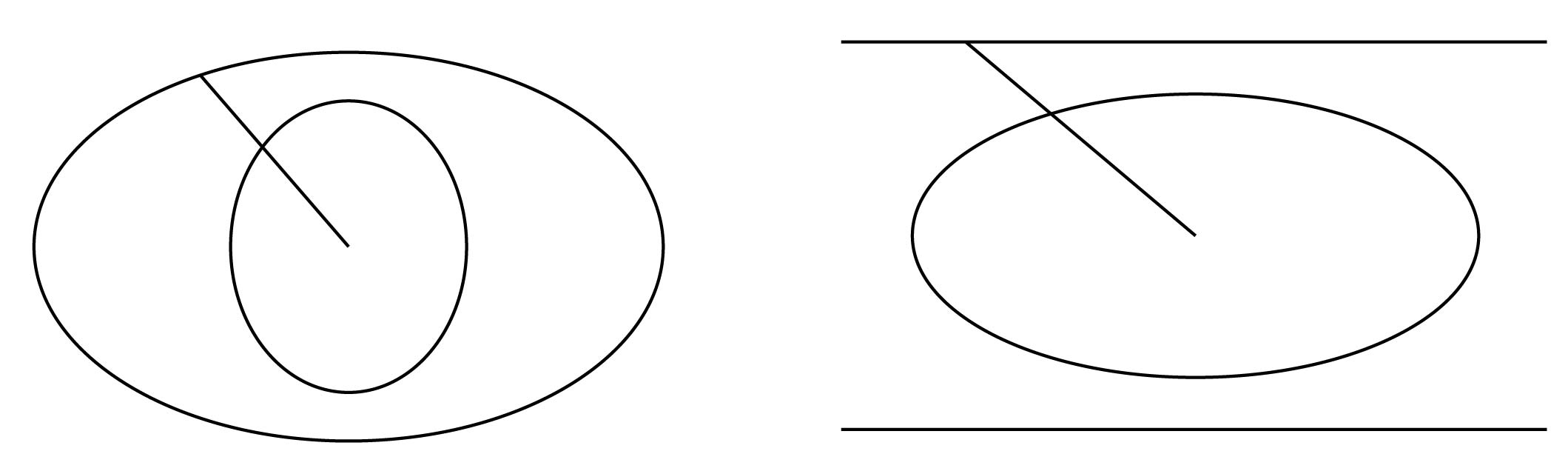}}
\centerline{Figure 1. Correspondence between two spaces of constant curvature.}
\medskip

Consider the list of the most classical integrable problems from mechanics, those considered for example in Arnold's textbook \cite{Arn}. There is the central force problem, proposed and integrated by Newton in Proposition XLI of the {\it Principia}, the motion of a free rigid body, proposed and integrated by Euler in 1765 (see \cite{Cay}, p.\ 567), the problem of a heavy rigid body with rotational symmetry, attached at a point of its axis, integrated by Lagrange in 1788. Later in the textbook there appears a fourth classical problem, the two fixed centers problem, proposed and integrated by Euler in 1766 (see \cite{Eu1}, \cite{Eu2}) and then a fifth one, the problem of the geodesics on an ellipsoid,  which was integrated by Jacobi in 1838 (see \cite{Ja1}).

The first three problems are integrated after a reduction of the rotational symmetry. The same is true for the last one in the special case of an axisymmetric ellipsoid. These integrations may be considered as the most elementary ones.

The fourth problem is an illuminating application of the model we present here. We get something that was missing in all the classical works, namely, a geometrical interpretation of the quadratic first integral discovered by Euler. It is enough to introduce the spherical two fixed centers problem, and to check that it corresponds with Euler's classical problem by central projection. The pull-back of a quadratic first integral is a quadratic first integral. We have an energy in each problem, so we get two linearly independent quadratic first integrals (see \cite{Alb}).

We cannot directly deal with the fifth problem, since the ellipsoid is not projectively flat. However, Kn\"orrer \cite{Kno} discovered that the Gauss map sends all the orbits of this problem onto orbits of another classical integrable problem, proposed and integrated by Neumann in 1859, where a particle moves on a sphere under a quadratic potential. The sphere is projectively flat. Our model applies and shows its full effectiveness: the integrability of this problem extends to an arbitrary dimension. We infer the integrability by just exhibiting a second centered quadric in correspondence with the sphere through central projection (see \cite{Al2}).

Note that the interesting ideas in \cite{MaT} and \cite{Ta1}, which allow one to deal directly  with the geodesics on the $n$-dimensional ellipsoid, were published a year before Lundmark's licentiate thesis \cite{Lu0}. They develop ideas presented by Levi-Civita as a generalization of Appell's remarks (see \cite{L-C}, \cite{Th1}, \cite{Ben}). Appell's paper was subsequently neglected and forgotten, while the more advanced ideas in Levi-Civita's paper survived.

The class of integrable problems we are dealing with has been distinguished for a long time, starting maybe in 1846 with Liouville \cite{Lio}, and was the object of many further studies. All the first integrals are quadratic. Arnold's textbook thus contains some classical problems integrated by mere observation of the symmetries and of the corresponding linear first integrals, and other problems whose integrability is deduced from a pair of quadratic first integrals. Of course there exist less elementary cases of integration. The Kowalevski top is a typical example with a quartic first integral.

Our point is not to extend these previous studies but to present  the most elementary point of view on the most elementary class of systems. These systems are not only quasi-bi-Hamiltonian. They are {\it naturally} Hamiltonian in two different ways, and we show the two configuration spaces, the two kinetic energies, the two Hamiltonians. This pushes remarkable studies developed after Lundmark's claim a step forward in the direction of simplicity (see \cite{CS1}, \cite{Ben}, \cite{Tsi}).

 There is indeed a yet more elementary way to reach  the figure of the two centered quadrics. One can avoid the use of connections by starting from the remark that a force field that is positively homogeneous of degree $-3$ may always be reduced by one degree of freedom (see \cite{Al2}).

The general study of the situation with two quadrics will be developed in a subsequent work. Here we focus on the above question of characterizing the pairs $(\mu, g)$ associated to a given force field. The study of their relation to a centered quadric readily shows the importance of some tools which only appear after long developments in the traditional studies.

To look for a $g$ satisfying (i) we should look for a quadratic first integral of the system whose leading (quadratic) term possesses certain algebraic properties. To fully understand these properties, we introduce the homogeneous form of an arbitrary polynomial first integral. Its leading term
is encoded by a tensor with Young tableau symmetry, the tableau being rectangular with two rows. The role of these tensors was observed only recently in a similar context (see \cite{MMS}). We decided to give a rather extended presentation of the theory of tensors with Young tableau symmetry, being unable to locate an appropriate reference, in particular concerning the implicit characterization of these tensors, which gives the fastest access to the properties which are needed here.

Therefore {\bf Sections 2 to 6} of this paper constitute a rather long analysis which can be presented as a study of the tensors with Young tableau symmetry, together with what should perhaps be understood as the most elementary motivation of this theory: the study of the polynomial first integrals of the free motion on an affine space. We also discuss the first integrals which are polynomial or rational in the velocity variable, while depending {\it a priori} arbitrarily on the position variable.

{\bf Sections 7 to 10} concern the properties of the leading term of an energy integral obtained from a $(\mu,g,U)$ with properties (i) and (ii). The algebraic properties of such a term are studied by Lundmark, who called them of cofactor type. In our presentation such a leading term is associated to a $g$ which is constrained by Beltrami's theorem. To reach some interesting examples related to the $n$-body problem, some degenerate cases should be included, which gives us a pretext to present a slightly extended version of Beltrami's theorem, together with a proof which looks quite different from the known proofs of this classical theorem.

\bigskip
\centerline{\bf 2. Dynamical homogeneous coordinates and free motion}
\bigskip

Recall that an affine space $A$ of dimension $n$ may always be seen as an affine
hyperplane in a vector space $V$ of dimension $n+1$. Intrinsically, $V$ is the dual of the
vector space of the affine functions on $A$. Our notation always assumes that $A\subset V$, in such a way that the points of $A$ are denoted as vectors of $V$. We denote by $\vec A$ the direction of $A$, i.e.\ the vector hyperplane parallel to $A$. The ground field of our vector and affine spaces is implicitly assumed to be $\R$.

Consider a system of affine coordinates $q_1,\dots, q_n$ on $A$. Each $q_i$ may be considered as  a linear form on $V$. The system of {\it homogeneous coordinates} $q_0,q_1,\dots, q_n$ is obtained by adding the linear form $q_0$ such that the equation of $A$ in $V$ is $q_0=1$. Choose an integer $k$. A function $f$, defined on a subset of $A$, is lifted to the function:
$$F:(q_0,\dots,q_n)\longmapsto q_0^kf\Bigl({q_1\over q_0},\dots,{q_n\over q_0}\Bigr),\eqno(2.1)$$
which is a homogeneous function of degree $k$ defined on a subset of $V$. When $f$ is a polynomial of degree $p$, we choose $k=p$ and the lifted function $F$ is a homogeneous polynomial of degree $p$.

We propose an extension of this standard construction. We extend this classical lift to functions defined on a subset of the tangent space $TA=A\times \vec A$. To  such a function $g(q_1,\dots, q_n;\dot q_1,\dots,\dot q_n)$ we associate the function:
$$G:(q_0,q_1,\dots,q_n;\dot q_0,\dot q_1,\dots,\dot q_n)\mapsto g\Bigl({q_1\over q_0},\dots,{q_n\over q_0};q_0\dot q_1-q_1\dot q_0,\dots,q_0\dot q_n-q_n\dot q_0\Bigr),\eqno(2.2)$$
defined on a subset of $V\times V$. Note that $g$ is the restriction of $G$ to the subspace $q_0=1$, $\dot q_0=0$.  

Consider the free motion on $A$, defined by the ordinary differential equation $\ddot q=0$. The linear first integrals of this motion (where linear means linear in the velocity vector $\dot q$) are linear combinations of $$\dot q_1,\,\dot q_2,\dots, \dot q_n,\, q_1\dot q_2-q_2\dot q_1,\dots,q_{n-1}\dot q_n-q_n\dot q_{n-1}.$$ The lifted forms of these elementary first integrals are respectively
$$q_0\dot q_1-q_1\dot q_0,\,q_0\dot q_2-q_2\dot q_0,\dots, q_0\dot q_n-q_n\dot q_0,\,q_1\dot q_2-q_2\dot q_1,\dots,q_{n-1}\dot q_n-q_n\dot q_{n-1}.$$ Let $q=(q_0,q_1,\dots, q_n)\in V$. The above list is the list of the coordinates of $q\wedge\dot q$. Three remarkable properties appear. First, the lifted forms are still polynomial expressions. Second, there is just one kind of formula in the new list, instead of two in the old list. Third, all the lifted functions are antisymmetric by exchange of position and velocity vectors.

{\bf Remark.} Consider natural variations on Formula $(2.2)$. We can multiply this expression by $q_0^k$. We can replace all the expressions $q_0\dot q_i-q_i\dot q_0$ by $q_0^m(q_0\dot q_i-q_i\dot q_0)$. If $(k,m)=(0,-2)$ then $g$ is lifted to a well-defined function on $T{\cal P}(V)$, the tangent space of the projective space. This choice of $(k,m)$ looks quite familiar, but this is not our choice here. The above antisymmetry property is only true if $k+m=0$. Our choice $(k,m)=(0,0)$ is the only one having also the following property. 

{\bf 2.1. Lemma.} Let $q_a\in A$, $q_b\in A$, $q_a\neq q_b$. Assume that a real function $g$ is defined on $\Omega\subset A\times\vec A$ and satisfies  for all $\gamma>0$ $$g\bigl(q_a, \gamma(q_b-q_a)\bigl)=g\bigl(q_b,\gamma(q_b-q_a)\bigl).\eqno(2.3)$$ The subset $\Omega$ is only assumed to contain all the points involved in the equation. Lift the function $g$ to a function $G$ according to Formula $(2.2)$. Then for any $\mu>0$, $\nu>0$ and $\gamma>0$, we have, after setting $Q_a=\mu q_a$, $Q_b=\nu q_b$,
$$G\bigl(Q_a, \gamma(Q_b-Q_a)\bigr)=G\bigl(Q_b,\gamma(Q_b-Q_a)\bigr),\eqno(2.4)$$
$$G(Q_a, Q_b)=G(Q_b,-Q_a).\eqno(2.5)$$

{\sl Proof.} By construction, if $G$ is well-defined at $(q,v)\in V\times V$, then for any $\lambda>0$ and any $\gamma\in\R$, $G(\lambda q,\lambda^{-1}v)=G(q,v)=G(q,v+\gamma q)$. Thus $$G\bigl(Q_a, \gamma(Q_b-Q_a)\bigr)=G\bigl(\mu q_a,\gamma(\nu q_b-\mu q_a)\bigr)=G\bigl(\mu q_a,\gamma(\nu q_b-\nu q_a)\bigr)=$$
$$=G\bigl(q_a,\gamma\mu\nu( q_b- q_a)\bigr)=g\bigl(q_a,\gamma\mu\nu( q_b- q_a)\bigr)=g\bigl(q_b,\gamma\mu\nu( q_b- q_a)\bigr).$$
This last equality is Hypothesis $(2.3)$. By the same computation in reverse order, we get $G\bigl(Q_b, \gamma(Q_b-Q_a)\bigr)$ and the conclusion $(2.4)$. Finally, we get
$$G( Q_a, Q_b)=G(Q_a, Q_b-Q_a)=G( Q_b, Q_b-Q_a)=G(Q_b,-Q_a)$$
by using $(2.4)$ in the particular case $\gamma=1$.\qed

{\bf 2.2. Remark.}  Assume the weaker hypothesis that $(2.3)$ is true for any $\gamma$ belonging to an open interval containing $1$. Then the computation in the proof shows that $(2.4)$ is still true if $\gamma\mu\nu$ belongs to this interval,  and that $(2.5)$ is still true if $\mu\nu$ belongs to this interval.

Hypothesis $(2.3)$ is clearly a property of a first integral $g$ of the differential system $\ddot q=0$.
We gave to this property the form $(2.5)$, a kind of antisymmetry property of the lifted function $G$. The relevance of this new remark appears in the following deduction. Hereafter, ``polynomial'', or ``rational'', implicitly means polynomial, or rational, with real coefficients.

{\bf 2.3. Proposition.} Let $U\subset A$ be a convex open set of the affine space $A$. Let $g: U\times \vec A\to\R$, $(q,\dot q)\mapsto g(q,\dot q)$, be a first integral of the differential system $\ddot q=0$. Suppose that for any fixed $q\in U$, $\dot q\mapsto g(q,\dot q)$ is a polynomial function of $\dot q$. Then $g$ is a polynomial function of $(q,\dot q)$. Furthermore, the lifted function $G$ obtained from $g$ by expression $(2.2)$ is a polynomial function of $(q,\dot q)\in V\times V$.

{\sl Proof.} The lifted function $G$ is a polynomial function of the second variable when the first is fixed in $\hat U=\{\mu q, \mu>0, q\in U\}$.  According to Lemma 2.1, $G$ satisfies $G(Q,Q')=G(Q',-Q)$, for any $Q$ and $Q'$ in $\hat U$.  So, $G(Q,Q')$ is also a polynomial in the first variable when the second is fixed. According to a result by Palais \cite{Pal}, $G$ is a polynomial function. Then, also is $g$, by restriction.\qed

This result improves a result by Nijenhuis \cite{Nij}, the main part of which being also proved in Thompson \cite{Thn} by a different but as elegant argument. Here we do not assume the continuity of $g$, and we do not even assume that the degree of $\dot q\mapsto g(q,\dot q)$ is bounded when $q$ varies. 

Similar results may be proved about the first integrals of $\ddot q=0$ which are rational in velocity instead of polynomial (see recent advances on related problems in \cite{Koz}). Painlev\'e (\cite{Pai}, p.\ 94) stated such a result. His argument starts with a first integral which is somewhere meromorphic in $(q,\dot q)$ and rational in $\dot q$. Using the property of a first integral, Painlev\'e excludes the appearance of a non-polar singularity in the analytic continuation of such a fraction. He then uses Appell's central projection in a crucial way to show that the first integral is rational. He concludes that the first integral is the quotient of two polynomial first integrals of $\ddot q=0$. We adapt here our Proposition 2.3 and propose a statement which may replace some of Painlev\'e's arguments. We suppose that $g$ is locally analytic on two disjoint convex open sets.

{\bf 2.4. Proposition.} Let $q_a\in A$, $q_b\in A$, $q_a\neq q_b$. Let $\Omega_a$ and $\Omega_b$ be two disjoint convex open sets in $A\times\vec A$,  respectively neighborhoods of $(q_a,q_b-q_a)$ and of $(q_b,q_b-q_a)$. Let $\Omega=\Omega_a\cup\Omega_b$, and let $U_a$, $U_b$ and $U$ be the respective projections of $\Omega_a$, $\Omega_b$ and $\Omega$ by the canonical projection $A\times\vec A\to A$. Let $g: \Omega \to\R$ be a locally analytic function. Suppose that for any $(q,v,\gamma)\in \Omega_a\times\R$ sufficiently close to $(q_a,q_b-q_a,1)$, we have $g(q, v)=g(q+\gamma v,v)$. Suppose moreover that for any $q\in U$, there is a rational function in $v$ that coincides with $v\mapsto g(q,v)$ on $(\{q\}\times \vec A)\cap \Omega$. Then there are two relatively prime polynomials $r$ and $s$ on $A\times\vec A$, first integrals of $\ddot q=0$, such that $g$ coincide on $\Omega$ with the fraction $r/s$.

{\sl Proof.} We follow the same argument as in the previous proof. We lift $g$ to a $G$ which is defined around $(q_a, q_b-q_a)$ in $V\times V$, and consequently around $(q_a,q_b)$, according to Formula $(2.2)$. In the same way, $G$ is defined around $(q_b, q_b-q_a)$ in $V\times V$, and consequently around $(q_b,-q_a)$. By Remark 2.2, if $(Q,Q')$ is close enough to $(q_a,q_b)$, then $G(Q,Q')=G(Q',-Q)$.  So  $G(Q,Q')$ is rational in $Q'$ when $Q$ is fixed and close to $q_a$, and rational in $Q$ when $Q'$ is fixed and close to $q_b$. According to Theorem 5, p.\ 201, in Bochner \& Martin \cite{BocM}, $G$ coincides, near $(q_a,q_b)$, with a rational function in the variable $(Q,Q')$. Of course $G$ is also rational near $(q_b,-q_a)$ due to $G(Q,Q')=G(Q',-Q)$. We claim that $G$ is the same fraction in both neighborhoods. Indeed, $g(q,  v)=g(q+\gamma v,v)$ for $\gamma$ around $1$ also implies that $g(q, v)=g(q+\epsilon v,v)$ for $\epsilon$ around 0, as
$g(q+\epsilon v,v)=g(q+\epsilon v+(\gamma-\epsilon)v,v)=g(q+\gamma v,v)$. This identity concerns a rational function, and can thus be extended to arbitrary values of $\epsilon$. By taking $\epsilon$ close to $1$ we see that this rational function coincides with $g$ in the neighborhood of $(q_b,-q_a)$.

Thus we have a first integral of the form $r/s$, where $r$ and $s$ are relatively prime polynomials in $(q,v)$ with real coefficients. Then $\dot r$ and $\dot s$, considered as polynomials obtained by formal derivation of $r$ and $s$ with respect to time along the flow of $\ddot q=0$,  satisfy $r\dot s=s\dot r$. Then $r$ divides $\dot r$, and $s$ divides $\dot s$. As a consequence, the flow of $\ddot q=0$ leaves invariant the hypersurface $r=0$, even in the complex domain. But along any orbit $t\mapsto q_0+tv_0$, $r$ is a polynomial in $t$, well-defined in the complex domain. If not constant, $r(t)$ has zeros, which contradicts the invariance of $r=0$. Then $r$ is always constant: $r$ is a polynomial first integral of $\ddot q=0$. The same is true for $s$.\qed

The previous results insist on the first integrals of $\ddot q=0$ which are polynomial in $(q,\dot q)$. They furthermore show that such a first integral $g$ has a polynomial lift $G: V\times V\to \R$ which satisfies, for any $\lambda\neq 0$ and any $\gamma\in\R$, 
$$G(\lambda q,\lambda^{-1}v)=G(q,v)=G(q,v+\gamma q)=G(q+\gamma v,v).\eqno(2.6)$$
These relations imply a cascade of properties. Let us state a rather unexpected converse to the end of Proposition 2.3, which shows the tight relation between the two objects of this section, the free motion $\ddot q=0$ and the dynamical homogeneous coordinates.

{\bf 2.5. Theorem.} Let $U\subset A$ be a convex open set and $g: U\times \vec A\to \R$. If $g$ is lifted through expression $(2.2)$ to a polynomial function $G$ of all of its variables, then $g(q,\dot q)$ is a polynomial first integral of $\ddot q=0$.

This statement will be naturally proved by simply recalling the famous properties of polynomials $G(q,v)$ that satisfy $(2.6)$ and that are homogeneous of degree $b$ in $v$ (and then also in $q$). The theorem is an easy consequence of the following algebraic proposition, that we will include in a general framework in the next three sections.

{\bf 2.6. Proposition.} Let $b\in\N$ and   $G:V\times V\to\R$ be a polynomial in two vector variables, satisfying for any $\lambda\in \R$ and any $\gamma\in\R$:
$$G(\lambda q,v)=G(q,\lambda v)=\lambda^bG(q,v),\quad G(q,v)=G(q,v+\gamma q).$$
Then $G$ also satisfies $G(q+\gamma v,v)=G(q,v)$ and $G(q,v)=(-1)^bG(v,q)$.

\bigskip
\centerline{\bf 3. Multilinear forms with Young tableau symmetry.}
\centerline{\bf Multi-alternate case.}
\bigskip

In this section we prepare a proof of Proposition 2.6 by collecting the basic properties of the vector space generated by the family of multilinear forms
$$\varphi_1\wedge\varphi_2\wedge\cdots\wedge\varphi_{j_1}\otimes\varphi_1\wedge\varphi_2\wedge\cdots\wedge\varphi_{j_2}\otimes\cdots\otimes\varphi_1\wedge\varphi_2\wedge\cdots\wedge\varphi_{j_c},\eqno(3.1)$$
where $c\geq 1$ is a fixed integer, $j_1,\dots, j_c$ are fixed integers satisfying $j_1\geq j_2\geq\cdots\geq j_c\geq 1$, where $V$ is a finite dimensional real vector space, and where $(\varphi_1,\dots,\varphi_{j_1})$ varies in $(V^*)^{j_1}$.

In this formula the exterior product $\wedge$ has priority on the general tensor product $\otimes$. We should read for example $\mu\wedge\nu\otimes\mu=(\mu\wedge\nu)\otimes\mu=(\mu\otimes\nu-\nu\otimes\mu)\otimes\mu=\mu\otimes\nu\otimes\mu-\nu\otimes\mu\otimes\mu$.
This family of $N$-linear forms, with $N=j_1+\cdots+j_c$, generates a subspace of
$$\bigwedge^{j_1}V^*\otimes\bigwedge^{j_2}V^*\otimes\cdots\otimes\bigwedge^{j_c}V^*$$
which we wish to characterize in several ways. If for example $c=1$, this subspace is the full $\bigwedge^{j_1} V^*$, consisting of all the antisymmetric $j_1$-linear forms. If  $j_1=j_2=\cdots=j_c=1$, this subspace is not the full $\bigotimes^c V^*$, but the subspace of all the symmetric $c$-linear forms. Consider now the next case, namely $c=2$, $j_1=2$, $j_2=1$. The subspace generated by all the 3-linear forms  $\mu\wedge\nu\otimes\mu$, with $(\mu,\nu)\in V^*\times V^*$, is the space of the 3-linear forms $\Phi\in\bigwedge^2V^*\otimes V^*$ such that $\Phi(x,y;z)+\Phi(y,z;x)+\Phi(z,x;y)=0$, for any $(x,y,z)\in V^3$. We will now present the general case.

A {\it Young tableau} is just a table that organizes the integer parameters $j_1,\dots,j_c$ in a convenient way. The {\it shape} of a Young tableau is given by the number $c \geq 1$ of the columns and by the list $[j_1,\dots,j_c]\in \N^c$, $j_1\geq j_2\geq\cdots\geq j_r\geq 1$, of the heights of these columns. A Young tableau is given by its shape and by a numbering of its boxes. For example, this is a Young tableau with shape $[4,3,3,2,2]$.
$$\begin{Young}
  1  &  5 &  8 & 11 & 13\cr
     2  &  6  & 9 & 12 & 14   \cr
     3  &  7 & 10    \cr
    4           \cr
\end{Young}$$
\centerline{Table 3.1}

We associate the $k$-th variable of the forms we study to the $k$-th box in the
tableau, according to the numbering. In this section, we implicitly adopt the numbering from top to bottom, then from left to right, as in Table 3.1. We associate to the tableau two operators $S$ and $A$, the {\it Young symmetrizers}, acting on $\bigotimes^N V^*$.

The operator $S$  symmetrizes a multilinear form $\Phi\in \bigotimes^N V^*$ in the variables corresponding to the boxes of the first row {\it and} symmetrizes in the variables corresponding to the boxes of the second row {\it and} etc. We can write $S$ as the composition of $j_1$ commuting operators: $S=S_1S_2\cdots S_{j_1}$, where $S_j$ is the symmetrization operator corresponding to the boxes of the $j$-th row. For example if the tableau has shape $[2,2]$, and is numbered as we said, $(S\Phi)(x,y;z,t)=\Phi(x,y,z,t)+\Phi(z,y,x,t)+\Phi(x,t,z,y)+\Phi(z,t,x,y)$.

The operator $A$  antisymmetrizes a multilinear form $\Phi\in \bigotimes^N V^*$ in the
boxes of the first {\it column} and in the boxes of the second column,  etc. We can write $A$ as the composition of $c$ commuting operators: $A=A_1A_2\cdots A_c$, where $A_i$ is the antisymmetrization operator corresponding to the boxes of the $i$-th column. With the same $2\times 2$ tableau, $(A\Phi)(x,y;z,t)=\Phi(x,y,z,t)-\Phi(y,x,z,t)-\Phi(x,y,t,z)+\Phi(y,x,t,z)$.

To write correctly such relations, we memorize that the rows are associated to symmetry, the columns to antisymmetry, and we
work mentally with diagrams as
$$\matrix{\begin{Young}
$x$  &  $z$ \cr
$y$  &  $t$    \cr\end{Young}} - \matrix{\begin{Young}
$y$  &  $z$ \cr
$x$  &  $t$    \cr\end{Young}}- \matrix{\begin{Young}
$x$  &  $t$ \cr
$y$  &  $z$    \cr\end{Young}}+ \matrix{\begin{Young}
$y$  &  $t$ \cr
$x$  &  $z$    \cr\end{Young}}.$$
\centerline{Table 3.2}

{\bf 3.1. Proposition.} For any Young tableau, there is a nonzero $\lambda\in \N$ such that $A:\bigotimes^N V^*\to\bigotimes^N V^*$ and
$S:\bigotimes^N V^*\to\bigotimes^N V^*$ defined above satisfy $SASA=\lambda SA$ and $ASAS=\lambda AS$.

This result is due to Alfred Young \cite{Yo2}, p.\ 364. A simplified proof by J.\ v.\ Neumann is presented in \cite{VdW}, p.\ 192, in \cite{We1}, p.\ 363 or in \cite{We2}, p.\ 124. These proofs are purely combinatorial and do not use any theoretical background. They concern the first identity $SASA=\lambda SA$. We obtain the second identity $ASAS=\lambda AS$ by transposition of the first, noticing that ${}^t\! S$ acts on $\bigotimes^N V$ as $S$ acts on $\bigotimes^NV^*$, i.e.\ by the same process of symmetrization, and that the same is true for $A$. 

{\bf 3.2. Proposition.} The subspace of $\bigwedge^{j_1}V^*\otimes\bigwedge^{j_2}V^*\otimes\cdots\otimes\bigwedge^{j_c}V^*$ generated by all the elements $(3.1)$ is the image of $AS$.
 
{\sl Proof.} In the case $j_1=\cdots=j_c=1$ and consequently  also in the general case, the image of $S$ is generated by the elements $\varphi_1\otimes\cdots\otimes
\varphi_{j_1}\otimes\varphi_1\otimes\cdots\otimes\varphi_{j_2}\otimes\cdots\otimes \varphi_1\otimes\cdots\otimes\varphi_{j_c}$, i.e.\
elements having the linear form $\varphi_i$ repeated along the $i$-th row. Applying $A$ this gives the result.\qed

{\bf 3.3. Proposition.} An $N$-linear form $\Phi\in {\rm Im} AS$ is zero if
for any $(x_1,\dots,x_{j_1})$ in $V^{j_1}$,
$$\Phi(x_1,x_2,\dots,x_{j_1};x_1,x_2,\dots,x_{j_2};\dots;x_1,x_2,\dots,x_{j_c})=0.\eqno(3.2)$$

{\sl Proof.} By the same argument as in the previous proof, Equation $(3.2)$
implies that $\langle \Phi,x\rangle=0$ for any $x\in{\rm Im} {}^t\!S$. This means $S\Phi=0$. But there exists a $\Psi$ such that $\Phi=AS\Psi$. So $0=ASAS\Psi=\lambda
AS\Psi=\lambda\Phi$.\qed

{\bf 3.4. Theorem.} We consider a Young tableau with shape $[j_1,\dots,j_c]$ and boxes numbered vertically as in Table 3.1, in such a way that $s_k=j_1+\cdots+j_{k-1}+1$ is the numbering of the $k$-th box of the first row. Let $N=j_1+\cdots +j_c$. For any $(m,n)$, $1\leq m<n\leq N$, we denote by ${\cal T}_m^n$ the transposition operator acting on $\bigotimes^NV^*$. It exchanges the variables $m$ and $n$ of an $N$-linear form $\Phi\in\bigotimes^NV^*$. An $N$-linear form $\Phi\in\bigotimes^NV^*$ is an element of ${\rm Im}\, AS$ if and only if

(i) it is antisymmetric in each column: for any column $k$, for any $(m,n)$, $s_k\leq m<n<
s_{k+1}$, ${\cal T}_m^n\Phi=-\Phi$,

(ii) it satisfies $r-1$ identities: for any column $k$, $1\leq k<c$,
$$\Phi-{\cal T}_{s_k}^{s_{k+1}}\Phi-{\cal T}_{s_k+1}^{s_{k+1}}\Phi-\cdots-{\cal T}_{s_k+j_k-1}^{s_{k+1}}\Phi=0.$$

{\bf Remarks on identities (ii)}. Identities (ii) appear in various contexts and particular cases.
$$\begin{Young}
    &  $*$ &  $s_3$ &  & \cr
       &  $*$  &  &  &    \cr
       &  $*$ &     \cr
               \cr
\end{Young}$$
\centerline{Table 3.3. A  set of four boxes involved in an identity (ii).}

They are mostly known as the ``algebraic Bianchi identities'' satisfied by the Riemann curvature tensor $R$. In index notation, they read $R_{ijkl}+R_{jkil}+R_{kijl}=0$.  In the same context, identities (i) are simply the antisymmetries $R_{ijkl}=-R_{jikl}$, $R_{ijkl}=-R_{ijlk}$. The ``if'' part of the theorem thus tells us that the quadrilinear form $R$ belongs to the image of $AS$, for the Young tableau with shape $[2,2]$.

$${\begin{Young}1&3\cr 2&4\cr\end{Young}}$$
\centerline{Table 3.4. Young tableau for the Riemann tensor.}

The proof of this ``if'' part is ``not trivially obvious'', already in the $[2,2]$ case, as is commented at the top of p.\ 144 of \cite{PeR}. A proof in this case is given in \cite{Aga}. We will propose a general proof of Theorem 3.4 in Section 6.

{\bf 3.5.  Example of an (easy) application.} Consider the space ${\cal B}\subset\bigotimes^4 V^*$ of the quadrilinear forms
$\phi$ such that, for any
$(x,y,z,t)\in V^4$,
$\phi(x,y,z,t)=-\phi(y,x,z,t)$, $\phi(x,y,z,t)=-\phi(x,y,t,z)$ and  $\phi(x,y,z,t)=\phi(z,t,x,y)$. Suppose $\phi\in{\cal
B}$ and
$\phi(x,y,x,y)=0$ for any $(x,y)\in V^2$. Then $\phi\in\bigwedge^4V^*$, i.e.\ $\phi$ is completely antisymmetric.

{\sl Proof.} Consider the projector $B:{\cal B}\to {\cal B}$, $\phi\mapsto\psi$ where $\psi$ is defined by
$\psi(x,y,z,t)=\phi(x,y,z,t)+\phi(y,z,x,t)+\phi(z,x,y,t)$. As $B^2=3B$, ${\cal B}=\ker B\oplus {\rm Im} B$.
We check directly that $\psi$ is antisymmetric so ${\rm Im} B=\bigwedge^4 V^*$. By Theorem 3.4, $\ker B={\rm Im} AS$
for the Young tableau $Y=[2,2]$ numbered vertically. We decompose $\phi=\phi_Y+\psi/3$ according to the decomposition of
${\cal B}$. Finally we write $\phi(x,y,x,y)=\phi_Y(x,y,x,y)=0$. Thus $\phi_Y=0$ by Proposition 3.3.\qed

The end of this section is borrowed from Towber \cite{Tow}. Identities (ii) are sometimes referred to as ``Fock cyclic identities'' (see \cite{Foc}, part 2). When $(3.1)$ is $\varphi_0\wedge\varphi_1\wedge\cdots\wedge\varphi_{n}\otimes\varphi_0\wedge\varphi_1\wedge\cdots\wedge\varphi_{n}$, where $(\varphi_0,\dots,\varphi_n)$ is a basis of $V^*$, they reduce to a particular case of identities published by Sylvester in 1851, concerning the product of two similar determinants. The general case of Sylvester's identities also extends to the general ${\rm Im }AS$, in the form of the next proposition, which is Corollary 1, p.\ 423 of \cite{Tow}.  Another identity on the curvature tensor, namely the block symmetry $R_{ijkl}=R_{klij}$, a well-known  consequence of the algebraic Bianchi identity, is also a particular case of this proposition.

{\bf 3.6. Proposition.} With the same notation as in Theorem 3.4, assume that $\Phi\subset{\rm Im} AS$. Extend the transposition operator ${\cal T}$ to simultaneous transpositions as follows: if $H$ and $K$ are two disjoint subsets of boxes of the Young tableau with same cardinality (we write $\#H=\#K$), ${\cal T}_K^H$ exchanges the block of variables $K$ and the block of variables $H$ while respecting the order of variables of the Young tableau.
Let $I$ be any column, and $J$ any subset of a posterior column of the Young tableau. Then
$$\Phi=\sum_{\scriptstyle K\subset I\atop\scriptstyle \#K=\#J}{\cal T}_{K}^{J}\Phi.$$

{\bf Example.} If we choose  $I$ as the first column below, and $J$ as a pair of boxes in the third column, the above sum will have 6 terms corresponding to all the choices of 2 boxes among the 4 boxes of the first column.
$$\begin{Young}
 $*$   &   &  $*$ &  & \cr
    $*$   &    &  $*$ &  &    \cr
   $*$    &   &     \cr
         $*$      \cr
\end{Young}$$
\centerline{Table 3.5. A  set of boxes involved in a generalized identity.}

Note that even when $\#J=1$ this new identity is more general than the identity (ii) as it also applies to nonadjacent columns.

\bigskip
\centerline{\bf 4. Multilinear forms with Young tableau symmetry.}
\nobreak
\centerline{\bf Multi-symmetric case.}
\nobreak
\bigskip
The image of the composition $SA$ of the Young symmetrizers $A$ and $S$ possesses essentially the same properties as the image of $AS$, studied in the previous section. A simple description by generators as $(3.1)$ is missing, but Theorem 4.1 below is exactly similar to Theorem 3.4, except for the signs in the identities.

To get an example of tensor in the image of $SA$, it is enough to apply $S$ on the  Riemann tensor. The result is the {\it symmetrized Riemann tensor} (see e.g.\ \cite{Syn}, p.\ 54), a tensor that contains exactly the information of the Riemann tensor.

To state Theorem 4.1, we shall change our convention about shape and numbering of a Young tableau, in order to keep the traditional association of rows to symmetry and of columns to antisymmetry. We shall give the shape of the Young tableau by the number $r\geq 1$ of the rows and the list $(i_1,\dots,i_r)\in \N^r$, $i_1\geq i_2\geq\cdots\geq i_r\geq 1$, of the lengths of these rows.
Our preferred numbering is the horizontal one, as described in Table 4.1, which presents a Young tableau different from the one  in Table 3.1, but with same shape. This shape was denoted by $[4,3,3,2,2]$ and is now denoted by $(5,5,3,1)$.

$$\begin{Young}
  1  &  2 &  3 & 4 & 5\cr
     6  &  7  & 8 & 9 & 10   \cr
     11  &  12 & 13     \cr
     14           \cr
\end{Young}$$
\centerline{Table 4.1}
\goodbreak

{\bf 4.1. Theorem.} We consider a Young tableau with shape $(i_1,\dots,i_r)$ and boxes numbered horizontally as in Table 4.1, in such a way that $s_k=i_1+\cdots+i_{k-1}+1$ is the numbering of the $k$-th box of the first column.
Let $N=i_1+\cdots +i_r$. For any $(m,n)$, $1\leq m<n\leq N$, we denote by
${\cal T}_m^n$ the transposition operator acting on
$\bigotimes^NV^*$. An $N$-linear form
$\Phi\in
\bigotimes^NV^*$ is an element of ${\rm Im}\, SA$ if and only if

(i) it is symmetric in each row: for any row $k$, for any $(m,n)$, $s_k\leq m<n<
s_{k+1}$, ${\cal T}_m^n\Phi=\Phi$,

(ii) it satisfies $r-1$ identities: for any row $k$, $1\leq k<r$,
$$\Phi+{\cal T}_{s_k}^{s_{k+1}}\Phi+{\cal T}_{s_k+1}^{s_{k+1}}\Phi+\cdots+{\cal T}_{s_k+i_k-1}^{s_{k+1}}\Phi=0.$$

{\bf Remarks on identities (ii).} As their counterpart  in Theorem 3.4,  identities (ii) are studied by Towber, who presents the multi-symmetric case (see \cite{Tow}, Definition {2}.{4}) after the multi-alternate one, showing the astonishing analogy of both cases. Towber proves his results for an arbitrary ring $R$ with unit, while we work only with the ground field $\R$. Of course our algebraic statements immediately pass to $\Q$ or $\C$. 
$$\begin{Young}
     &    &    &   &  \cr
     $*$  &  $*$  & $*$ & $*$ & $*$   \cr
     $s_3$  &    &       \cr
                 \cr
\end{Young}$$
\centerline{Table 4.2. A set of six boxes involved in an identity (ii).}

\bigskip
\centerline{\bf 5. Young tableau symmetry and free motion.}
\bigskip

A polynomial first integral of the free motion $\ddot q=0$ may be decomposed into homogeneous components in velocity $\dot q$, and each component is a first integral. Such a homogeneous term, after the particular homogeneization in position $q$ described in Section 2, is uniquely associated to a tensor with Young tableau symmetry, the tableau being a rectangle with two rows (compare \cite{MMS}). Proposition 2.6, which we proposed as the explanation of the curious Theorem 2.5, is a standard property of such tensors. We give here an elementary deduction of these facts, based on the general theory presented in the previous sections. We could also prove Proposition 2.6 without this preparatory material. We could use Propositions {2}.{1} to {2}.{3} of \cite{Tow}, and thus avoid any mention of the Young symmetrizers $A$ and $S$. But some other identities would remain obscure.

{\bf 5.1. Definition.} For any $b\in\N$, we call ${\cal P}^{b,b}(V)$ the space of polynomials $G:
V\times V\to\R$, $(q,v)\mapsto G(q,v)$ which (i) are homogeneous of degree $b$ in each vector variable and (ii) satisfy
$G(q,v+\gamma q)=G(q,v)$ for any $(q,v,\gamma)\in V\times V\times\R$.
 
We now express these properties using the ``polar form'' $G_{\cal S}$ of $G$. This is the unique $2b$-linear form  on $V$, symmetric in the first $b$
arguments, symmetric in the last $b$ arguments, such that:
$$G(q,v)=G_{\cal S}(q,\dots,q;v,\dots,v).\eqno(5.1)$$
The polar form $G_{\cal S}$ is obtained, as is well-known, by a repeated differentiation of the polynomial $G$. Condition
(i) fixes the number $2b$ of the arguments. Condition (ii) may be written  $dG(q,v+\gamma q)/d\gamma=0$, which gives
$$G_{\cal S}(q,\dots,q;q,v,\dots,v)=0,\qquad\hbox{for any } (q,v)\in V\times V.\eqno(5.2)$$

$$\begin{Young}$q$&$q$&$q$&$q$&$q$\cr $q$&$v$&$v$&$v$&$v$\cr\end{Young}$$
\centerline{Table 5.1. Diagram for Condition (ii).}

{\bf 5.2.  Proposition.} Let $b\in\N$ and $G_{\cal S}$ be a $2b$-linear form, symmetric in the first $b$
variables, symmetric in the last $b$ variables. Let $G$ be the polynomial given by expression
$(5.1)$. Let $S$ and $A$ be the Young symmetrizers associated to the Young tableau with shape $(b,b)$, numbered from left to right, then top to bottom. Then:
 $$G\in {\cal P}^{b,b}(V)\quad\Longleftrightarrow\quad(5.2) \hbox{ is satisfied }\quad\Longleftrightarrow\quad
 G_{\cal S}\in {\rm Im}SA.$$

{\sl Proof.} Condition $(5.2)$ is equivalent to: the symmetrization of $G_{\cal S}$ in its
$b+1$ first arguments gives zero. This is identity (ii) of Theorem 4.1.\qed

{\bf 5.3. Proposition.} A polynomial $G$ is in ${\cal P}^{b,b}(V)$ if and only if there exists a $2b$-linear form
$G_{\cal A}$ which is, for any
$i$, $1\leq i\leq b$, antisymmetric by  exchange of the $2i-1$-th and $2i$-th variable, such that
$$G(q,v)=G_{\cal A}(q,v;q,v;\dots;q,v).\eqno(5.3)$$
Consequently a $G\in{\cal P}^{b,b}(V)$ satisfies $G(q,v)=(-1)^bG(v,q)$.

{\sl Proof.} If $G_{\cal A}$ is given, then $G$ is defined by $(5.3)$ as its symmetric part, due to repeated $q$ and repeated $v$. In formulas, $(b!)^2G(q,v)=(S\,G_{A})(q,v;q,v;\dots;q,v)$, where $S$ is the Young symmetrizer. The polar form $G_{\cal S}$ of $G$ is then $(b!)^{-2}S\,G_{\cal A}$, which is clearly an element of ${\rm Im}SA$. We apply Proposition 5.2 and conclude that $G\in{\cal P}^{b,b}(V)$. To prove now the only if part, we start from $G$, deduce using Proposition 5.2 that $G_{\cal S}\in {\rm Im}SA$. Clearly $AG_{\cal S}$ is, up to a factor, a correct (but not the unique) choice for $G_{\cal A}$, due to Proposition 3.1. \qed

{\bf 5.4.  Definition.} We call ${\cal P}_{\cal S}^{b,b}(V)$ the space of $2b$-linear $G_{\cal S}$ forms satisfying
the symmetry condition of Proposition 5.2 and the condition $(5.2)$. We call  ${\cal P}_{\cal A}^{b,b}(V)$ the space of $2b$-linear forms $G_{\cal A}$ satisfying the antisymmetry condition of Proposition 5.3, and the 

{\bf 5.5. Identities (ii).} For any $i$, $1\leq i\leq b-1$, the antisymmetrization of $G_{\cal A}$ in the
$2i-1$-th,
$2i$-th and $2i+1$-th variables gives zero.

So, ${\cal P}_{\cal S}^{b,b}(V)={\rm Im}SA$, $S$ and $A$ being the Young symmetrizers associated to the Young tableau
$(b,b)$ numbered horizontally. And according to Theorem 3.4, ${\cal P}_{\cal A}^{b,b}(V)={\rm Im} AS$, where $A$ and
$S$ are the Young symmetrizers associated to the Young tableau with same shape but numbered vertically.  This may be stated as
follows.

{\bf 5.6.  Proposition.} If $G\in {\cal P}^{b,b}(V)$, there exists a unique $G_{\cal A}\in {\cal P}^{b,b}_{\cal A}(V)$
satisfying $(5.3)$. The $2b$-linear form $G_{\cal A}$ defines a unique $b$-linear form on $\bigwedge^2 V$,
which is symmetric and defines in turn uniquely a polynomial
$G_{\cal B}:\bigwedge^2V\to\R$, homogeneous of degree
$b$, such that $G_{\cal B}(q\wedge v)=G(q,v)$. The polynomial $G_{\cal B}$ is determined by its values on the decomposable bivectors.

{\sl Proof.} The first statement is clear: the expression of the unique $G_{\cal A}$ is given at the end of the previous proof. By Proposition 3.2, such a $G_{\cal A}$ is a linear combination of terms
$(\xi_1\wedge\xi_2)\otimes\cdots\otimes(\xi_1\wedge\xi_2)$, with $(\xi_1,\xi_2)\in (V^*)^2$, which  is a
linear combination of terms $\omega\otimes\cdots\otimes\omega$ with $\omega\in\bigwedge^2V^*=(\bigwedge^2V)^*$. This is
the symmetric $b$-linear form announced. Consider now two such tensors with same value on the decomposable bivectors. Proposition 3.3 proves that their difference is zero.\qed

Clearly $\R\oplus {\cal P}^{1,1}(V)\oplus {\cal P}^{2,2}(V)\oplus\cdots$
is an algebra for the multiplication of polynomials.
The dimension of ${\cal P}^{b,b}(V)$ is $$\frac{n(n+1)^2(n+2)^2\cdots
(n+b-1)^2(n+b)}{b!(b+1)!}$$ if $\dim V=n+1$. This is an application of  a general result proved in 1954 in \cite{FRT} and presented in \cite{PeR} . Let us fix $b=5$ for an example of tableau. One computes the numerator of this fraction by multiplying together all the numbers in Table 5.2, and the denominator by multiplying the numbers in Table 5.3, which gives the ``hook length'' of the Young tableau.

$$\begin{Younglong}$n+1$&$n+2$&$n+3$&$n+4$&$n+5$\cr $n$&$n+1$&$n+2$&$n+3$&$n+4$\cr\end{Younglong}$$
\centerline{Table 5.2}

$$\begin{Young}$6$&$5$&$4$&$3$&$2$\cr $5$&$4$&$3$&$2$&$1$\cr\end{Young}$$
\centerline{Table 5.3}

As a result about the dimension of the space of the first integrals of the free motion, this number was obtained without mention to a Young tableau symmetry, in \cite{KaL}, and rederived in many works. The result in the interesting particular case $b=2$ was derived earlier in \cite{Ths}.

Through $G_{\cal B}$ the polynomial $G$ may be interpreted as a polynomial on the Grassmannian of vectorial 2-planes in
$V$. This interpretation generalizes to $p$-planes, $p\geq 2$ (see \cite{FuH} or \cite{Tow}, part 3). More significant to us, $G$ appears as
the most general polynomial in $q\wedge v$. In \cite{Al1} we called $q\wedge v$, in this dynamical context, the {\it projective impulse}. We have just established that a polynomial first integral of the free motion, i.e.\ the motion with constant projective impulse, is simply a ``polynomial in the projective impulse'', or a ``polynomial on the decomposable bivectors''.

\bigskip
\centerline{\bf 6. Characterization of Young tableau symmetry. The proofs.}
\bigskip

Our previous section provides a new application of the theory of multilinear forms with Young tableau symmetry, which is probably among the most elementary ones, although, as we just observed, this application does not appear as distinct from the theory of polynomial functions on the Grassmannian of 2-planes. We tried to give an easy access to this theory. We will conclude this elementary presentation by giving some proofs that we could not find anywhere. But first, we give a proof of  the ``only if'' part of Theorem 3.4, which was proved by Young (see \cite{Yo2}, \S 10), together with the ``only if'' part of Theorem 4.1, which is similar.

{\sl Proof of the ``only if'' part of Theorem 3.4.}  Consider the linear operator $B_k$ such that $B_k\Phi$ is the left-hand side of condition (ii) in this theorem. Consider the linear operator $A_k$, which antisymmetrizes with respect to the boxes in the $k$-th column. Then the composition $B_kA_k$ is simply the antisymmetrization with respect to $j_k+1$ boxes: the $j_k$ boxes of the $k$-th column and the box at the top of the $k+1$-th column. Our hypothesis is that $\Phi=AS\Psi$ for some $\Psi$. Suppose first that $j_{k+1}=1$, which means that there is only one box in the $k+1$-th column.
Then $S\Psi$ is symmetric by exchange of both boxes at the top of the $k$-th and the $k+1$-th columns. Then $B_kA_kS\Psi=0$, because we antisymmetrize a partially symmetric object. The required identity $B_kAS\Psi=0$ follows, as here, by the hypothesis $j_{k+1}=1$, $A=A_1\cdots A_{k-1}A_k$, where $A_j$ commutes with $B_k$ if $j<k$. If now $j_{k+1}\geq 2$, we form $\Lambda=S\Psi$, expand $A_{k+1}\Lambda$ and apply $B_kA_k$ to each term. Each gives zero, as there is a partial symmetry in each term of the expansion, corresponding to the different row symmetries of $\Lambda$. We conclude as above, observing that $A_{k+2},\dots, A_c$ also commute with $B_k$.\qed

The proofs of the ``if'' parts are more difficult to locate in the literature. We already mentioned the notes by Penrose and Rindler and by Agacy. A related, but {\it a priori} weaker statement may be found in Young \cite{Yo2} and Towber \cite{Tow}. They prove that all the operators in the algebra of the permutation group which annihilate ${\rm Im} AS$ (respectively ${\rm Im} SA$), are generated by identities (i) and (ii). Young stated this in his Section 10 and proved it in his Section 12. Towber strengthens Young's result in the multi-alternate case, in his Theorem {4}.{1}, and his Corollary 2, p.\ 446. 

Let us begin our elementary proof of this ``if'' part. The conclusion of the lemma below is the case of Proposition 3.6 that we briefly indicated at the very end of Section 3.

{\bf 6.1. Lemma.} If $\Phi$ satisfies identities (i) and (ii) of Theorem 3.4, then

(iii) $\Phi$ satisfies these $r(r-1)/2$  identities: for any $(j,k)$, $1\leq k< j\leq c$,
$$\Phi-{\cal T}_{s_k}^{s_{j}}\Phi-{\cal T}_{s_k+1}^{s_{j}}\Phi-\cdots-{\cal T}_{s_k+j_k-1}^{s_{j}}\Phi=0.$$

 The argument of the proof is that two consecutive identities (ii) combine and give an identity (iii) with $j=k+2$, and so on. We will not reproduce the careful expansion, but will rather refer to \cite{Tow}, p.\ 426, where Towber extends this kind of transitivity to his block identities, that we explained in our Proposition 3.6.

{\bf 6.2. Lemma.} Consider again a Young tableau with shape $[j_1,\dots,j_c]$. Consider a $\Phi\in\bigotimes^NV^*$, $N=j_1+\cdots+j_c$, satisfying hypotheses (i) and (ii) of
Theorem 3.4. Let $p\in V$ be an arbitrary vector. Then for any integer $\ell$, $1\leq \ell\leq c$, the multilinear form $\Phi'\in\bigotimes^{N-\ell}V^*$,
obtained by contracting
$p$ in each top box of the last $\ell$ columns, satisfies Relations (i) and (ii) of Theorem 3.4 for the Young sub-tableau with shape
$[j_1,\dots,j_{c-\ell},j_{c-\ell+1}-1,\dots,j_{c'}-1]$, where $c'$ is the first number $\geq c-\ell$ for which $j_{c'+1}\leq 1$.
$$\begin{Young}&&$p$&$p$&$p$\cr&&&&\cr&&\cr\cr\end{Young}$$
\centerline{Table 6.1. Illustration of Lemma 6.2.}

{\sl Proof.} Obviously $\Phi'$ satisfies identity (i). 
If $k<c-\ell$, identity (ii) is the same equation for $\Phi$ and $\Phi'$.
If $k\geq c-\ell$, identity (ii) for $\Phi$ can be written as well using the second top box in column $k+1$ instead of the top box (if there is no second top box, then no identity is needed).  We have two sub-cases. If $k=c-\ell$, we have the same equation for $\Phi$ and $\Phi'$. If $k> c-\ell$, there is one term less for $\Phi'$, but this term is zero, because $p$ is repeated in column $k+1$.\qed

{\sl Proof of the ``if'' part of Theorem 3.4.} Let ${\cal B}$ be the subspace of $\bigotimes^NV^*$ defined by (i) and (ii). Note that (i) means ${\cal B}\subset{\rm Im}A$. The ``only if'' part, which we just proved, is ${\rm Im}AS\subset{\cal B}$.  We know by Proposition 3.1 that $S: {\rm Im}AS\to{\rm Im}SA$ is bijective. It is thus
sufficient to prove that $S:{\cal B}\to {\rm Im}SA$ is injective. So we want to prove that if
 $\Phi\in{\cal B}$ is such that $S\Phi=0$, then $\Phi=0$.
The hypothesis $S\Phi=0$ is equivalent to: 
$$\Phi(q_1,\dots,q_{j_1};q_1,\dots,q_{j_2};\dots;q_1,\dots,q_{j_c})=0\quad \hbox{for any } (q_1,\dots,q_{j_1})\in
V^{j_1}.\eqno(6.1)$$ For any $\Phi$ satisfying (i) and (ii) we want to prove that $(6.1)\Rightarrow\Phi=0$.

Let us set
$$X(p)=\Phi(q_1,\dots,q_{j_1};\dots;q_1,\dots,q_{j_{c-1}};p,q_2,\dots,q_{j_c}).\eqno(6.2)$$
and $X_i=X(q_i)$. Hypothesis $(6.1)$ is then $X_1=0$. We will deduce that $X(p)=0$ for any $p$. This part of the proof can be represented by the following diagram.
$$\matrix{\begin{Young}$q_1$&$q_1$&$q_1$&$q_1$&$q_1$\cr$q_2$&$q_2$&$q_2$&$q_2$&$q_2$\cr $q_3$&$q_3$&$q_3$\cr$q_4$\cr\end{Young}}=0\qquad\Longrightarrow\qquad
X(p)=\matrix{\begin{Young}$q_1$&$q_1$&$q_1$&$q_1$&$p$\cr$q_2$&$q_2$&$q_2$&$q_2$&$q_2$\cr $q_3$&$q_3$&$q_3$\cr$q_4$\cr\end{Young}}=0$$
\centerline{Table 6.2}

We denote by $\partial_i X_i$ the derivative of $X_i$ with respect to the variable $q_i$, and compute this derivative in the direction of a variable $p$:
$$
	  \partial_i X_i \cdot p = \sum_{\ell=1}^c Y_i^\ell(p)+X(p),\eqno(6.3)
$$
where $Y_1^c(p)=0$, where $Y_i^\ell(p)=0$ if $i>j_\ell$, and where
$$Y_i^\ell(p)=\Phi(\dots;q_1,\dots,q_{j_{\ell-1}};q_1,\dots,q_{i-1},p,q_{i+1},\dots,q_{j_\ell};\dots;q_i,q_2,\dots, q_{j_c})\eqno(6.4)$$
if $1\leq i\leq j_\ell$, $(i,\ell)\neq (1,c)$. Note that for $i=2,\dots,j_c$, $X_i=0$ and $Y_i^c(p)=-X(p)$.
The identity (iii) of Lemma 6.1, concerning column number $\ell<c$ and column number $c$ of the Young tableau, gives $Y_1^\ell(p)+\cdots+Y_{j_\ell}^\ell(p)=X(p)$. Using this identity to deal with the $c-1$ first columns on the right-hand side of the system
$$\EQM{\partial_1X_1\cdot p&=Y_1^1(p)+\cdots+Y_1^{c-1}(p)+X(p),\cr
\partial_2X_2\cdot p&=Y_2^1(p)+\cdots+Y_2^{c-1}(p)+Y_2^c(p)+X(p),\cr&\vdots\cr
\partial_{j_1}X_{j_1}\cdot p&=Y_{j_1}^1(p)+\cdots+Y_{j_1}^{c-1}(p)+Y_{j_1}^c(p)+X(p),}\eqno(6.5)$$
we get
$$(\partial_1 X_1+\partial_2 X_2+\cdots+\partial_{j_1} X_{j_1})\cdot p=\bigl(c-1-(j_c-1)+j_1\bigr)X(p),\eqno(6.6)$$
where the factor $c+j_1-j_c$ is strictly positive.
We will prove by induction on $n$ that $X_n=X(q_n)=0$ for any $n$, $1\leq n\leq j_1$. By Formula $(6.6)$ this will imply that $X(p)=0$.

We know that $X_1=0$. For any $n=2,\dots, j_1$, we consider the sum
$$
	\sum_{i=1}^{n-1}\partial_i X_i \cdot q_n,
$$
which can be computed following the same procedure as in $(6.6)$. For that, we need the additional property 
$$
	Y_i^{\ell}(q_j)=0, \quad 
	\hbox{for }\ell=1,\ldots,c-1,\quad j\leq j_\ell 
	\quad\hbox{and}\quad j\neq i,
$$
which follows from the repetition of $q_j$ in column $\ell$. This implies, for $\ell=1,\ldots,c-1$, that
$$
	\sum_{i=1}^{n-1} Y_i^\ell(q_n)
	=\left\{\begin{array}{ll}
		0 & \hbox{if } n\leq j_\ell \\
		\sum_{i=1}^{j_\ell} Y_i^\ell(q_n)=X(q_n) & \hbox{if } n>j_\ell
	\end{array}
	\right.
$$
For $\ell=c$,
$$
	\sum_{i=1}^{n-1} Y_i^c(q_n)=\sum_{i=2}^{\min(n-1,j_c)} Y_i^c(q_n)
	=-\big(\min(n-1,j_c)-1\big)X(q_n).
$$
Hence, $k_n$ being the number of integer numbers $\ell$ such that $1\leq \ell\leq c-1$ and that $j_\ell<n$, we get
$$
	\sum_{i=1}^{n-1}\partial_i X_i \cdot q_n
	= \Bigl(k_n-\bigl(\min(n-1,j_c)-1\bigr)+n-1\Bigr) X_n.
$$
The integer on the right-hand side is always positive, being $k_n+1$ in the worst case $n-1\leq j_c$. This formula proves by induction $X_n=0$ for $n=1,2,\dots,j_1$, as announced. From Formula $(6.6)$ we deduce that $X(p)=0$ for any vector $p\in V$. The implication represented in Table 6.2 is proved.

We observe now that $X(p)=0$ is Equation $(6.1)$ written for the $N-1$-linear form $\Phi'$ obtained by contracting
$p\in V$ in $\Phi$, at the top of the last column. This form satisfies (i) and (ii) by Lemma 6.2. We prove
our assertion $(6.1)\Rightarrow \Phi=0$ by induction on the number of boxes $N$ of the Young tableau.\qed

{\sl Proof of Theorem 4.1.} The ``only if'' part is as in Theorem 3.4. Consider the ``if'' part,  consider a $\Phi$ satisfying the hypothesis and compute, for an arbitrary $(x_1,\dots x_r)\in V^r$,
$$
	(A\Phi)(\overbrace{x_1,\dots,x_1}^{i_1};
		\overbrace{x_2,\dots,x_2}^{i_2};\dots;
		\overbrace{x_r,\dots,x_r}^{i_r}).\eqno(6.7)
$$
In our example with shape $(5,5,3,1)$, this means that we ``fill up'' $A\Phi$  as shown in Table 6.3.
$$\begin{Young}
  $x_1$  &  $x_1$ & $x_1$  &  $x_1$  &  $x_1$ \cr
     $x_2$  &  $x_2$ &  $x_2$ &  $x_2$ &  $x_2$    \cr
$x_3$ &  $x_3$ &  $x_3$\cr
$x_4$\cr\end{Young}$$\centerline{Table 6.3}

The value of $A\Phi$ is a sum with signs of values of $\Phi$. In each term, the
positions of the variables in each column are exchanged. We consider any of these terms, e.g.\ the one corresponding to Table 6.4.
$$\begin{Young}
  $x_3$  &  $x_1$ & $x_3$  &  $x_1$  &  $x_2$ \cr
     $x_2$  &  $x_3$ &  $x_1$ &  $x_2$ &  $x_1$    \cr
$x_4$ &  $x_2$ &  $x_2$\cr
$x_1$\cr\end{Young}$$
\centerline{Table 6.4}

If we apply the last of the identities (ii), this term is changed into three terms, none of which possess $x_1$ in the
bottom row. Such operation would also work if there were already some $x_1$ in the third row. In this case several terms in
the identity (ii) would be equal. We would simply put them all on the same side of the equation, to get their
common value as the sum of the other terms, divided by an integer. Repeating this operation,
we create new terms where the $x_1$'s are higher and higher in the tableau, to finish with terms where they are all ``in their place'', in
the first row. When necessary we put $x_1$  in the first column using a transposition (i). We then place the $x_2$'s in
the second row, etc. Finally we express
$(6.7)$ as a sum where each term is a rational number times $\Phi(x_1,\dots,x_1;\dots;x_r,\dots,x_r)$. There is
a rational number $\mu$ which depends neither on $\Phi$ nor on the $x_i$'s, such that
$(A\Phi-\mu\Phi)(x_1,\dots,x_1;\dots;x_r,\dots,x_r)=0$. This is true for any $(x_1,\dots,x_r)\in V^r$ so
$S(A\Phi-\mu \Phi)=0$. As $\Phi$ satisfies (i), $S\Phi=i_1!i_2!\dots i_r!\Phi$.  
Thus $SA\Phi=\nu\Phi$ with $\nu=i_1!\dots i_r!\mu$. This equation implies that $\Phi\in{\rm Im} SA$ except in the bad case where $\nu=0$. But multiplying this equation by $SA$, using Proposition 3.1 and choosing any $\Phi\notin \ker SA$, we see that $\nu=\lambda\neq 0$.\qed

\bigskip
\centerline{\bf 7. Beyond the free motion on a flat space}
\bigskip

We said in the introduction that the left-hand side of $(1.3)$ has the form $D_{q'}q'$, where $D$ is a symmetric linear connection and $q'=dq/ds$.
The factor $\mu$ in the formula $D_{q'}q'=\mu^{-1}(\mu q')'$ comes from the change of time $(1.2)$, depending only on position. We can keep this new time and nearly recover the  simpler form $(1.1)$ by introducing a change $q\mapsto Q=\alpha q$, where $\alpha$ is a $q$-dependent positive factor. We expand
$${1\over \mu}\bigl(\mu q'\bigr)'={1\over \mu}\Bigl(\bigl({\mu\over\alpha}\bigr) Q'+\bigl({1\over\alpha}\bigr)'\mu Q\Bigr)'={1\over\alpha}Q''+\Bigl({1\over \mu}\bigl({\mu\over \alpha}\bigr)'+\bigl({1\over\alpha}\bigr)'\Bigr)Q'+\xi Q.\eqno(7.1)$$
We do not need the expression of the scalar quantity $\xi$, which will be soon related to a constraint. We make $\mu=\alpha^2$ in such a way that the middle term vanishes. Setting $\lambda=-\alpha\xi$, Equation $(1.3)$ becomes
$${d^2 Q\over ds^2}={f\over \alpha^3}+\lambda Q.\eqno(7.2)$$
This equation is the foundation of Appell's theory of central projection in mechanics. It may be interpreted as follows. 

\medskip
\centerline{\includegraphics [width=70mm]
{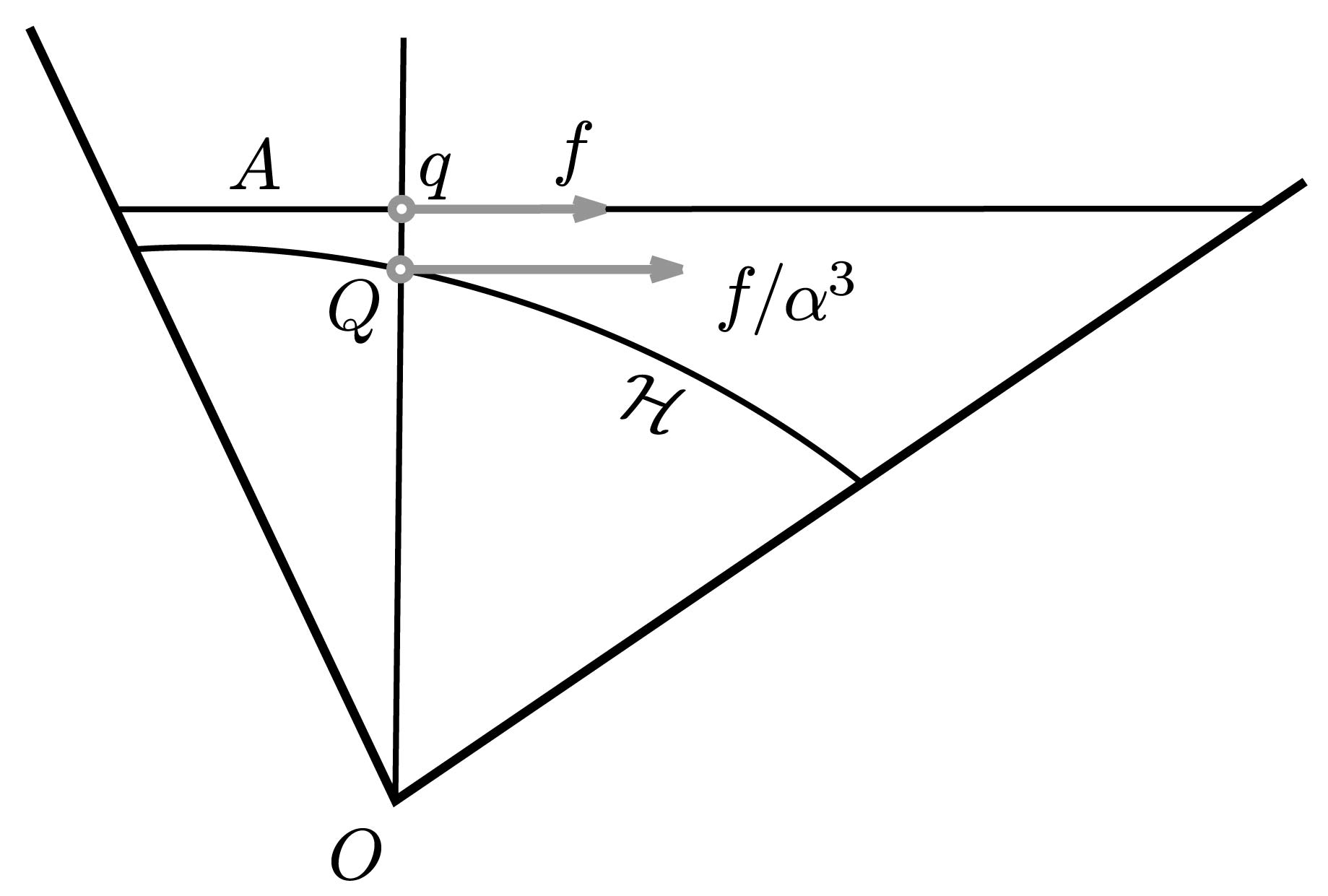}}
\centerline{Figure 2. Central projection on the screen ${\cal H}$.}
\medskip

The equation $Q=\alpha q$ refers to an origin, which should be placed out of the space of motion $A$. In other words, $A$ should be considered as an affine hyperplane of a vector space $V$. As $\alpha=\sqrt\mu$ only depends on the position $q$, the point $Q$ is constrained to move on a hypersurface ${\cal H}\subset V$, the image of ${\cal U}\subset A$ by the map $q\mapsto Q$, image which we called in \cite{Al1} a {\it screen}. The free motion on the screen is defined by
$${d^2 Q\over ds^2}=\lambda Q.\eqno(7.3)$$
The motions $(7.2)$ and $(7.3)$ are constrained in an unusual way. In the usual case of a particle constrained to move on a hypersurface, what replaces $\lambda Q$ is called the {\it reaction} and is {\it normal} to the hypersurface. Here the ``reaction'' $\lambda Q$ is {\it central} instead of normal. This central (also called ``radial'') direction is defined without using any Euclidean structure.

In the case of a centered spherical screen in a Euclidean space, radial and normal directions coincide: the motion $(7.2)$ is ``natural'', the geodesic motion $(7.3)$ is uniform along the great circles.  In the general case, a solution of $(7.3)$ describes a ``line'' on the screen, i.e.\  the intersection of the screen with a vectorial plane.

{\bf 7.1. Definition.}  Consider a finite dimensional real vector space $V$ and a hypersurface ${\cal H}\subset V\setminus\{0\}$, which at any of its points $q$ cuts transversally the vector line $[q]$. We call such a hypersurface a {\it screen}. At each $q\in{\cal H}$, consider the splitting $T_qV=T_{q}{\cal H}\oplus [q]$.  The {\it central connection} of ${\cal H}$ is the symmetric linear connection on ${\cal H}$ induced by the standard affine connection on $V$ and by this splitting.

The situation should again be compared to the case of a connection induced on a submanifold by the embedding in a Riemannian manifold. We adapt to our case the intuitive description of such a connection. The parallel transport defined by the central connection is decomposed as follows. A tangent vector $w$ should be first transported by infinitesimal parallelism in $V$, then centrally projected on the tangent space to ${\cal H}$. The corresponding differential equation is simply $w'|_{q}=\lambda q$, where $\lambda$ is, as in $(7.2)$ and $(7.3)$, the {\it ad hoc} multiplier whose value is forced by the tangency of the transported vector.
Now $(7.3)$ appears as the equation for the geodesics of the central connection.

The following lemma states the results of the above discussion and characterizes any of the discussed connections by specifying an invariant volume form.

{\bf 7.2. Lemma.} Let $A$ be an affine space of dimension $n\geq 1$, which we consider when needed as an affine hyperplane of a (real) vector space $V$ of dimension $n+1$. Call $\nu$ a constant volume form on $A$. Choose arbitrarily an open set  ${\cal U}\subset A$ and three smooth positive functions $\phi$, $\mu$ and $\alpha$ defined on ${\cal U}$.

(i)  There is a unique symmetric connection on ${\cal U}$ whose geodesics are carried by straight lines and such that
the volume form $\phi\nu$ is invariant by parallel transport.

(ii) There is a unique symmetric connection on ${\cal U}$ whose geodesics are given by the differential equation $(\mu q')'=0$. 

(iii) There is a unique symmetric connection on ${\cal U}$ which is sent by the map  ${\cal C}:q\mapsto \alpha q$ to the central connection of the hypersurface ${\cal C}({\cal U})\subset V$.

\noindent The connections (ii) and (iii) are equal when $\mu=\alpha^2$. The connections (i) and (iii) are equal when $\phi=\alpha^{n+1}$. To any connection characterized by (i), (ii) or (iii) corresponds a $\phi$, a $\mu$ and an $\alpha$ which are unique up to a constant positive factor.

{\sl Proof.} Statement (ii) was  explained in the introduction of this paper after Formula $(1.3)$. We should only recall that a symmetric connection is uniquely determined by its geodesics. Statement (iii) and the relation $\mu=\alpha^2$ were proved in the introduction of this section. 

Let us prove the existence part of (i) by showing that the connection (iii) with $\alpha=\root n+1 \of \phi$ has the required property. Let $\omega\in\bigwedge^{n+1} V^*$ be a constant volume form on $V$. The contracted product $q\lint\omega$ at each point $q\in{\cal H}={\cal C}({\cal U})$ defines a volume form on ${\cal H}$ which is clearly invariant by the parallel transport associated to the central connection. If $v_1,\dots, v_n$ form a base of $T_qA$, their standard $A$-volume is given by $q\wedge v_1\wedge \cdots\wedge v_n$, while their ${\cal H}$-volume is given by $(\alpha q)\wedge (\alpha v_1)\wedge\cdots\wedge (\alpha v_n)$. The ${\cal H}$-volume is then $\alpha^{n+1}\nu=\phi\nu$. So the connection preserves the required volume.

Let us prove the uniqueness in (i). According to the classical works \cite{Wey} and \cite{Eis}, the difference of  two geodesically equivalent symmetric connections is a 3-tensor $u_{ij}^{\phantom{k}k}=\xi_i\delta_j^k+\xi_j\delta_i^k$ where $\xi$ is a differential 1-form, and $\delta_k^i$ the Kronecker $\delta$. The difference  of the two induced linear connections on the volume bundle is the 1-form $u_{ik}^{\phantom{k}k}=(n+1)\xi_i$. As both connections preserve the same volume, this difference is zero, thus $\xi=0$ and $u=0$.\qed

{\bf 7.3. Remarks and definitions.} The covariant derivative operator $D$ associated to the connection (ii) is defined by the formula $D_XX=\mu^{-1}\partial_X(\mu X)$ for any vector field $X$ on ${\cal U}$, where $\partial _X$ is the standard derivation along the vector $X$.

The hypersurface ${\cal H}={\cal C}({\cal U})$ associated in (iii) to a symmetric connection will be called {\it the uniformizing screen} of this connection. As stated in the last part of the lemma this hypersurface is only unique up to homothety. Strictly speaking, we should not say {\it the} uniformizing screen but {\it a} uniformizing screen.

In dimension $n\geq 2$, a short computation shows that the connection determined by $\alpha$ is flat if and only if $1/\alpha$ is an affine function on ${\cal U}$. The uniformizing screen is then part of an affine hyperplane in $V$.

A connection on ${\cal U}\subset A$ whose geodesics are carried by straight lines (in other words whose {\it unparametrized geodesics}, also called {\it pre-geodesics}, are straight lines) is called {\it geodesically equivalent} to the affine connection.  A connection on an open set $\Omega$ of a manifold is called {\it projectively flat} if there exist an affine space $A$, an open set ${\cal U}\subset A$, and a diffeomorphism $\Omega\to{\cal U}$ which pushes forward the connection on a connection which is geodesically equivalent to the affine connection.

Consider a projectively flat connection. The definition associates to it an affine space $A$. If we have an invariant volume, we can use Lemma 7.2 to build a uniformizing screen ${\cal H}={\cal C}({\cal U})$. We have already discussed the question of the uniqueness of the uniformizing screen, but here this question is raised again, in a different way, since $A$ is not associated uniquely to the projectively flat connection.

We consider another affine space $A_1$ associated to the same connection. We pull back to $A$ the affine connection of $A_1$ by the convenient composition of the diffeomorphisms associated to $A$ and $A_1$. We get a flat connection on $A$ which is geodesically equivalent to the affine connection. As we just said, if $n\geq 2$, its uniformizing screen is part of an affine hyperplane in $V$. This is a realization of $A_1$. Now, we have in the same vector space $V$ two affine hyperplanes $A$ and $A_1$ and the screen ${\cal H}$ that we built from $A$. As central projections may be composed, the screen ${\cal H}$ may also be obtained from $A_1$ instead of $A$. This shows that if $n\geq 2$, ${\cal H}$ is, up to an arbitrary linear transformation, associated in a unique way to the projectively flat connection.

{\bf 7.4. Projective impulse.} Recall that, starting with a Newton system $(1.1)$, we first changed the time $t\mapsto s$ and then applied a central projection $q\mapsto Q$. The relation $\mu=\alpha^2$ between the parameters of these two transformations implies the fundamental identity $$q\wedge {dq\over dt}=Q\wedge {dQ\over ds},\eqno(7.4)$$ which shows the role of the decomposable bivector $q\wedge dq/dt$, which we called in \cite{Al1} the {\it projective impulse}, and which is constant along the orbits of $(7.3)$. The relation with the dynamical homogeneous coordinates described in Section 2 is clear: a function obtained by the lifting $(2.2)$ takes the same value on a dynamical data $(q,dq/dt)$ on $A$, and on the corresponding dynamical data $(Q,dQ/ds)$ on the screen ${\cal H}$.

{\bf 7.5. Invariant scalar products.} These remarks allow us to present a new proof of Beltrami's theorem (compare \cite{Bel},  \cite{Wey}, \cite{Mat}), which includes a rather original argument. The hypothesis of this theorem is that a scalar product, the pseudo-Riemannian structure, is conserved by the parallel transport associated to one of the connections described in Lemma 7.2. 

We will begin with the original dimension treated by Beltrami, a 2-dimensional surface. We will extend his original hypothesis. We will treat pseudo-Riemannian scalar products instead of only Riemannian. This is a known extension. We will also extend the hypothesis to the case where a degenerate scalar product is invariant by parallel transport.  In this new case we furthermore assume that an area form is preserved. Such a form is also preserved in the nondegenerate cases, as a consequence of the preserved nondegenerate scalar product (and orientability).

{\bf 7.6. Theorem.} Consider a connected open set ${\cal U}$ of an affine {\it plane} $A$, endowed with a connection which is geodesically equivalent to the affine connection. If a nondegenerate scalar product is invariant by parallel transport, then ${\cal U}$ is part of a pseudo-Riemannian surface of constant curvature. The uniformizing screen (see \S 7.3) is part of a nondegenerate quadric or of an affine plane.
If an area form and a rank one scalar product are invariant by parallel transport, then the degeneracy curves are parts of lines which either all intersect at a point or are all parallel. The uniformizing screen is part of a cylinder whose generating lines carry the direction of  degeneracy of the scalar product\footnote{By cylinder we mean a general cylinder: the cross section is not necessarily a conic section. The statement suggests to consider also the case where there is an invariant rank one scalar product (or an invariant linear form) but no invariant area form. Abdelghani Zeghib showed me that this weaker hypothesis would give a strictly weaker conclusion.}.

{\bf 7.7. Main argument of the proof.} The scalar product $\la ., .\ra$ is invariant by parallel transport, so $(q,q')\mapsto \la q',q'\ra$ is a quadratic first integral of the geodesic motion. Moreover, the connection satisfies property (i) of Lemma 7.2, which gives a function $\mu$ and property (ii). After a change of time $s\mapsto t$, defining the new velocity $\dot q=\mu q'$, we still have this quadratic first integral, now expressed as $(q,\dot q)\mapsto \mu^{-2}\la \dot q,\dot q\ra$, and now first integral of the free motion on $A$. Such an object is described by Proposition 2.3, Definition 5.1 and Proposition 5.6, and associated to a quadrilinear form with a $2\times 2$ Young tableau symmetry, which is also a bi-quadratic polynomial $\lk: V^2\to\R$, $(q,v)\mapsto \lk(q,v)$. Lemma 7.2 also gives a function $\alpha=\sqrt\mu$, a uniformizing screen ${\cal H}$ and a map ${\cal U}\to {\cal H}$, $q\mapsto  Q$. By $(7.4)$ and as $\lk$ may be seen as a quadratic form on $q\wedge\dot q=Q\wedge Q'$, we have $\lk(q,\dot q)=\lk(Q,Q')$. The quadratic form $Q'\mapsto \lk(Q,Q')$ is invariant by parallel transport on ${\cal H}$, since the quadratic form $q'\mapsto \la q',q'\ra$ is invariant by parallel transport on ${\cal U}$, and the respective quadratic forms and parallel transports correspond to each other through central projection.

From now on, we focus on ${\cal H}$ and denote by $q$ instead of $Q$ a generic point on ${\cal H}$. The parallel transport of a vector $w\in T_q{\cal H}$ along a path with tangent velocity $dq/ds=q'\in T_q{\cal H}$ is defined by the equation
$w'|_q=\lambda q$, where $\lambda$ is a real multiplier. We use the
antisymmetric polarization $\lk_{\cal A}$ of $\lk$, which is such that $\lk(q,w)=\lk_{\cal A}(q,w;q,w)$.  We compute
$$0=\frac{ d}{ds}\lk(q,w)=\frac{ d}{ds}\lk_{\cal A}(q,w;q,w)=2\lk_{\cal A}(q,w;q',w).$$
We arrive at a compatibility condition between ${\cal H}$ and $\lk_{\cal A}$:
$$\hbox{for any } q\in{\cal H},\; u\in T_q{\cal H},\; w\in T_q{\cal H},\quad \lk_{\cal A}(q,w;u,w)=0.\eqno(7.5)$$
We claim that $(7.5)$ implies: for any $q\in{\cal H}$, $(u,v,w)\in (T_q{\cal H})^3$, $\lk_{\cal A}(q,v;u,w)=0$.
Indeed, on $T_q{\cal H}$, $\lk_{\cal A}(q,.;.,.)$ vanishes on repeated $w$. Thus it is antisymmetric in positions 2 and 4. Being also antisymmetric in positions 3 and 4, this tensor is antisymmetric in 2, 3, 4. As it also satisfies the Bianchi identity in 2, 3, 4, it vanishes\footnote{This argument may be seen as the simplest example of reasoning on Young tableau symmetry: according to Lemma 6.2 we can apply Proposition 3.3 to the case of the trilinear form $\lk_{\cal A}(q,.;.,.)$.  The next simplest example would be 3.5, which we will use in the proof of Lemma 8.2.}.
Now $\lk_{\cal A}(q,v;u,w)$ depends on $q$ and $v$ through the bivector $q\wedge v$. So we can also write:
$$\hbox{for any } q\in{\cal H},\; v\in V,\; (u,w)\in (T_q{\cal H})^2,\quad \lk_{\cal
A}(q,v;u,w)=0.\eqno(7.6)$$
Consider a positively homogeneous function $h$ on the semi-cone generated by ${\cal U}$, such that $h(q)=1$ is an equation of the uniformizing screen ${\cal H}$. Thinking of $\lk_{\cal A}(q,v;.,.)$ as a 2-form in the missing arguments, the compatibility condition $(7.6)$ also reads:
$$\hbox{for any } q\in{\cal H}, v\in V,\quad dh|_q\wedge \lk_{\cal A}(q,v;.,.)=0.\eqno(7.7)$$

{\bf 7.8. End of the proof.} As $\dim V=3$, the (nonzero) symmetric linear map $\lk_{\cal B}: \bigwedge^2 V\to \bigwedge^2 V^*$ induced by $\lk_{\cal A}$ is conjugated, through the choice of a volume form on $V$, to a symmetric linear map $\lk_{\cal F}:V^*\to V$.  Condition $(7.7)$ can be written: for any $\xi\in V^*$ such that $\la \xi,q\ra=0$, $\la\lk_{\cal F}\xi,dh|_q\ra=0$. As $\lk_{\cal F}$ is symmetric, this reads $\lk_{\cal F} dh|_q\wedge q=0$. We will consider three cases according to the rank of $\lk_{\cal F}$.  If ${\rm rk} \lk_{\cal F}=3$, then the equation becomes $dh|_q\wedge \lk_{\cal F}^{-1}q=0$. So $h$ has the same level surfaces as $q\mapsto \la \lk_{\cal F}^{-1}q,q\ra$. We got the quadric and we easily check that the scalar product on it is induced by the quadratic form defining the quadric, giving a well-known model for a pseudo-Riemannian surface of constant curvature (see \cite{ONe}, p.\ 228). If ${\rm rk} \lk_{\cal F}=2$,  we write the compatibility equation in the form $\lk_{\cal F} dh|_q=\lambda(q) q$. If $q$ is not in the image of $\lk_{\cal F}$, the multiplier $\lambda(q)$ can only be zero. Consequently it vanishes everywhere. We see that this image is the direction of a planar screen. The  scalar product on this screen is induced by $\lk_{\cal F}$. It is nondegenerate and constant, or in other words, pseudo-Euclidean. Finally, all the cases with ${\rm rk} \lk_{\cal F}\geq 2$ give nondegenerate scalar products.

If ${\rm rk} \lk_{\cal F}=1$, there is a 2-dimensional kernel of $\lk_{\cal F}$ in $V^*$ corresponding, by contraction in a volume element, to a vectorial direction $[k]$ in $V$ such that $\lk_{\cal A}(k,.;.,.)=0$. In Vilms' terminology (see \cite{Vil}, p.\ 597), $k$ belongs to the {\it absolute kernel} of $\lk_{\cal B}$. We contract $k$ in $(7.7)$ and get, for all $q\in {\cal H}$, $v\in V$, $\la dh|_q,k\ra\lk_{\cal A}(q,v;.,,)=0$.  If $\la dh|_q,k\ra$ was nonzero at a point $q$, it would be nonzero on a neighborhood and $\lk_{\cal A}$ would be identically zero. But we assume that the scalar product is not trivial. Thus for all $q\in {\cal H}$, $\la dh|_q,k\ra=0$. The screen ${\cal H}$ is locally invariant by the translations of direction $k$. This direction, being the absolute kernel, is a degeneracy for the scalar product at any $q\in{\cal H}$.\qed

\bigskip
\centerline{\bf 8. On the linear maps preserving the decomposability of bivectors}
\bigskip

We are classifying the scalar products which are parallel for a projectively flat connection.  We recall that they are particular quadratic first integrals $\lk\in {\cal P}^{2,2}(V)$ of the free motion on an affine space. According to Proposition 5.6, $\lk$ has a polar form $\lk_{\cal A}\in {\cal P}_{\cal A}^{2,2}(V)$ with $2\times 2$ Young tableau symmetry, which in turn defines a symmetric linear map $\lk_{\cal B}: \bigwedge^2 V \to \bigwedge^2 V^*$. 

The  arguments in \S7.7 and \S7.8 classify such scalar products, degenerate or not,  when $\dim V=3$. Interestingly, if our aim was only to classify the parallel nondegenerate scalar products, we would not reach the conclusion sooner: the final argument is needed to establish that there are only degenerate scalar products in the third case.

What appeared indeed in this final argument is a simple algebraic criterium. The {\it absolute kernel} of $\lk_{\cal B}$  is defined as  the subspace $N=\{k\in V \hbox{ s.t.\ for any } u\in V,\;\lk_{\cal B}(k\wedge u)= 0\}$. {\it The parallel scalar product is nondegenerate if and only if the absolute kernel of $\lk_{\cal B}$ is trivial}.

The ``if'' part of this statement is the main technical difficulty when extending our version of Beltrami's theorem to higher dimension. We will overcome this difficulty here, by proving Proposition 8.1. We will then prove related algebraic results which are not easy to find in the literature in the form we will need.

{\bf 8.1. Proposition.} Let $V$ and $W$ be two real vector spaces with same finite dimension $\geq 3$. If a linear map $\lka:\bigwedge^2 V\to\bigwedge^2 W$ preserves decomposability, is not invertible but is such that ${}^t\lka$ has a trivial absolute kernel, then there is a nonzero $\phi\in W$ such that $\lka(\pi)\wedge \phi=0$ for any $\pi\in\bigwedge^2 V$.

{\bf Remark.} Here ${}^t\lka:\bigwedge^2W^*\to \bigwedge^2 V^*$ denotes the transpose of $\lka$. Its absolute kernel is a subspace of $W^*$. In our applications $W=V^*$ and ${}^t\lka=\lka$. We could use this symmetry to shorten a bit our proofs.

{\bf Basic tools.} We recall that a bivector $\pi\in\bigwedge^2 V$ is
called {\it decomposable} (or {\it simple}) if there exists a $(u,v)\in V^2$ such that
$\pi=u\wedge v$. We recall that $$\pi \hbox { is decomposable } \Longleftrightarrow\; \pi\wedge\pi=0\;\Longleftrightarrow\; {\rm rk}\,\pi\leq 2.$$ The linear map $\lk_{\cal B}$ satisfies $(7.7)$, which implies that it sends decomposable bivectors on decomposable 2-forms. In other words $\lk_{\cal B}$ preserves decomposability. Such linear maps, and especially those induced by an $\lk_{\cal A}\in {\cal P}_{\cal A}^{2,2}(V)$, were studied by Vilms (see \cite{Vil}). His statements are closely related with what we will prove here, but organized differently. We first recall the fundamental tools in his study.

We study the linear maps
$\lka:\bigwedge^2 V\to\bigwedge^2 W$ that send decomposable bivectors to decomposable bivectors, i.e.\
$$\hbox{for any } (u,v)\in V^2,\quad \lka(u\wedge v)\wedge \lka(u\wedge v)=0.\eqno(8.1)$$ The simplest way to obtain such an
$\lka$ is to take a linear map
$B: V\to W$, and form the unique operator
$B^{\wedge 2}:\bigwedge^2 V\to\bigwedge^2 W$ characterized by $$B^{\wedge 2}(u\wedge
v)=B(u)\wedge B(v).\eqno(8.2)$$ Not all the operators satisfying $(8.1)$ are obtained in this way. For example, if
$\dim V=3$,  $(8.1)$ is satisfied, but if ${\rm rk}\,\lka=2$, $\lka$ cannot be a $B^{\wedge 2}$.

{\bf 8.2.  Lemma.}  Let $S:\bigwedge^2 V\times\cdots\times\bigwedge^2V\to\R$,
$(\pi_1,\dots,\pi_p)\mapsto S(\pi_1,\dots,\pi_p)$ be a $p$-linear symmetric form. The following properties are equivalent

(i) $S(\sigma,\dots,\sigma)=0$ for any bivector $\sigma$ with ${\rm rk}\,\sigma\leq 2p-2$,

(ii) $S(\pi_1,\dots,\pi_{p-2},\pi,\pi)=0$ for any decomposable bivectors $\pi_1,\dots, \pi_{p-2}$, $\pi$, 

(iii) there exists a $T\in\bigwedge^{2p}V^*$ such that $S(\pi_1,\dots, \pi_p)=T(\pi_1\wedge\cdots\wedge \pi_p)$.

{\sl Proof.} Under condition (i) $S(\sigma,\dots,\sigma)=0$ if $\sigma=x_1 \pi_1+\cdots+x_{p-2}\pi_{p-2}+x\pi$, where $\pi_1,\dots, \pi_{p-2}$
and $\pi$ are decomposable and $(x_1,\dots,x_{p-2},x)\in\R^{p-1}$. As all the coefficients in the expansion of $S(\sigma,\dots,\sigma)$, the coefficient of $x_1\cdots x_{p-2}x^2$, being a positive integer times $S(\pi_1,\dots,\pi_{p-2},\pi,\pi)$, is zero.
This proves (i) $\Rightarrow$ (ii). All the coefficients in this expansion have a repeated bivector. This proves (ii)
$\Rightarrow$ (i). Consider the $2p$-linear form $t$ such that $t(u_1,\dots,u_{2p})=S(u_1\wedge
u_2,\dots,u_{2p-1}\wedge u_{2p})$. Assuming (ii),  $t(u_1,u_2,u_1,u_2,u_3,u_4,\dots,u_{2p-2})=0$. By \S 3.5,
$t$ is antisymmetric in its first four arguments. In the same way $t$ is antisymmetric in other similar sets of four arguments, and thus completely antisymmetric. This gives a $T$ with the required property on decomposable bivectors, and consequently, by linearity, on arbitrary bivectors.
This proves (ii) $\Rightarrow$ (iii). The remaining implication is standard.\qed

{\bf 8.3.  Lemma.}  For any $p\geq 2$, a linear map $\lka:\bigwedge^2V\to\bigwedge^2W$ satisfying $(8.1)$ induces a unique linear map $\lka^{(p)}:\bigwedge^{2p}V\to\bigwedge^{2p}W$ such that, for any $(\pi_1,\dots,\pi_p)\in(\bigwedge^2V)^p$, $\lka^{(p)}(\pi_1\wedge\cdots\wedge\pi_p)=\lka(\pi_1)\wedge\cdots\wedge \lka(\pi_p)$.

{\sl Proof.} Use (ii) $\Rightarrow$ (iii) in Lemma 8.2, with a $\bigwedge^{2p}W$-valued instead of a real
valued form, namely, the $p$-linear form $(\pi_1,\dots,\pi_p)\mapsto \lka(\pi_1)\wedge\cdots\wedge \lka(\pi_p)$.\qed

This lemma, in particular in the case $p=2$, appears as an essential argument in almost any proof about linear maps preserving decomposability of bivectors. We will now prove Proposition 8.1. We always assume that $\dim V=\dim W\geq 3$.

{\bf 8.4. Lemma.}  If a linear map $\lka:\bigwedge^2V \to\bigwedge^2 W$ preserves decomposability and is not invertible, then there is a nonzero decomposable bivector  in $\ker{}^t\lka$.

{\sl Proof.}  We distinguish two cases according to the parity of the dimension. We first consider $\dim V=\dim W=2m$. Let $\pi_1\in\ker \lka$. We can
find $\pi_2,\dots,\pi_m$ such that
$\pi_1\wedge\pi_2\wedge\cdots\wedge \pi_m\neq 0$. For example, if ${\rm rk}\,\pi_1=2k$, we make $\pi_2=\cdots=\pi_k=\pi_1$ and
choose convenient decomposable bivectors  $\pi_{k+1},\dots,\pi_m$. We see immediately from the definition of $\lka^{(m)}$ that $\lka^{(m)}=0$ on this nonzero $2m$-vector. As $\dim\bigwedge^{2m}V=1$, this simply means $\lka^{(m)}\equiv 0$.

Consider the smallest $k$ such that 
$\lka^{(k+1)}\equiv 0$. We can find a decomposable $2k$-vector $\eta\in{\rm Im} \lka^{(k)}$ among the images of the $2k$-vectors forming a standard base of $\bigwedge^{2k} V$. These images are not all  zero and are decomposable.  Due to $\lka^{(k+1)}\equiv 0$, the wedge product of $\eta$ with any bivector in ${\rm Im} \lka$ is zero. For any $l\in \N$, any $w_1,\dots, w_l$,  $\zeta=\eta\wedge w_1\wedge\cdots\wedge w_l$ has the same property. We can form a nonzero $\zeta$ as soon as $2k+l\leq \dim W$. We choose $l=\dim W-2(k+1)$, and fix a nonzero $\mu\in\bigwedge^{\dim W}W^*$. For any $\pi\in {\rm Im} \lka$, $0=\la\pi\wedge\zeta,\mu\ra=\la \pi,\zeta\lint\mu\ra$. The contracted product $\zeta\lint\mu$ is clearly a decomposable 2-form in $\ker {}^t\lka$.

Suppose now $\dim V=\dim W=2m+1$. Consider a first subcase where there exists a non-decomposable bivector $\pi_1\in\ker \lka$. Let $(e_0,\dots,e_{2m})$ be a base of $V$ such that
$\pi_1=e_0\wedge e_1+e_2\wedge e_3+\cdots$. It is easy to see that the elements of the standard base of $\bigwedge^{2m}
V$ all have the form $\pi_1\wedge\pi_2\wedge\cdots\wedge \pi_m$, with, for $i\geq 2$, $\pi_i=e_{i_1}\wedge e_{i_2}$,
$0\leq i_1<i_2\leq {2m}$. For example to get $e_1\wedge\cdots\wedge e_{2m}$, we take $\pi_2=e_1\wedge e_4$,
$\pi_3=e_5\wedge e_6$, etc. From this and the
definition of $\lka^{(m)}$, we deduce that $\lka^{(m)}\equiv 0$. We conclude exactly as in the even-dimensional case.

We continue the odd-dimensional case with the other subcase, with only decomposable bivectors in $\ker\lka$. Suppose $e_0\wedge e_1$ is such a bivector. Then $\lka^{(m)}$ is zero on all the $2m$-vectors of a standard base of $\bigwedge^{2m}V$, except maybe on $e_0\wedge f$ and $e_1\wedge f$, with $f=e_2\wedge\cdots\wedge e_{2m}$. If $\lka^{(m)}(e_0\wedge f)=\lka^{(m)}(e_1\wedge f)=0$, then $\lka^{(m)}\equiv 0$ and we conclude as in the previous cases. Suppose $\lka^{(m)}(e_1\wedge f)\neq 0$. Then $\rho=\lka^{(m-1)}(e_1\wedge e_2\wedge\cdots\wedge e_{2m-2})$ and $\eta=\lka(e_{2m-1}\wedge e_{2m})$
are nonzero and decomposable. Choose a nonzero $\kappa\in W$ such that  $\eta\wedge \kappa=0$.  Then $\zeta=\rho\wedge\kappa$ is nonzero and decomposable. We check that for any $\pi$ of the form $\lka(e_i\wedge e_j)$, and consequently for any $\pi\in{\rm Im}\lka$, we have $\zeta\wedge \pi=0$. We conclude as in the previous cases.\qed

{\sl Proof of Proposition 8.1.}  Lemma 8.4 gives a nonzero decomposable bivector $\xi\wedge\eta\in\ker{}^t\lka$. The pair $(\xi,\eta)\in (W^*)^2$ is such that, according to $(8.1)$, for all $(u,v)\in
V^2$,
$$0=(\xi\wedge \eta)\lint\bigl( \lka(u\wedge v)\wedge \lka(u\wedge v)\bigr)=2\bigl(\xi\lint \lka(u\wedge v)\bigr)\wedge \bigl(\eta\lint
\lka(u\wedge v)\bigr).$$
This is the collinearity of two elements of $W$. To see how this collinearity may occur, we take the value of
the right-hand side on $(\alpha,\beta)\in (W^*)^2$ and find
$$\la \lka(u\wedge v),\xi\wedge\alpha\ra\la \lka(u\wedge v),\eta\wedge\beta\ra=\la \lka(u\wedge v),\xi\wedge\beta\ra
\la \lka(u\wedge v),\eta\wedge\alpha\ra.$$
Here a polynomial in the coordinates of the vectors $u$, $v$, $\alpha$ and $\beta$
possesses two factorizations. The degrees are the
same: if the polynomial $\la \lka(u\wedge v),\xi\wedge\beta\ra$ divides the polynomial $\la \lka(u\wedge
v),\eta\wedge\beta\ra$, the quotient is a real number $\lambda$. Then $(\eta-\lambda\xi)\lint \lka(u\wedge v)=0$ for
all $(u,v)\in V^2$, which is excluded by the hypotheses of a trivial absolute kernel and $\xi\wedge\eta\neq 0$.

But $\la \lka(u\wedge v),\xi\wedge\beta\ra$ cannot divide the other factor, which does not contain the variable $\beta$. So it cannot be irreducible as a polynomial
in the coordinates of $u$, $v$, $\beta$. We claim that the factorization is
$$\la \lka(u\wedge v),\xi\wedge\beta\ra=\la\omega,u\wedge v\ra\la\phi,\beta\ra, \quad\hbox{with}\quad
\omega\in\bigwedge^2V^*,\quad\phi\in W.$$
The proof is as follows. As the product is homogeneous, the factors are homogeneous. The total degree in the coordinates of $\beta$ is one for the product, so it should be one
for a factor, zero for the other factor.  The same is true
for the coordinates of $u$ and of $v$. So the factorization is: linear times bilinear. But the expression is antisymmetric in $u$ and $v$, which consequently enter in the same factor.
So there exist $\omega$ and $\phi$ as claimed. We have $\xi\lint \lka(u\wedge v)=\la\omega,u\wedge v\ra\phi$. As the absolute kernel is trivial, this quantity is not identically zero. Contracting $\xi$ in $(8.1)$ gives the result $\lka(u\wedge v)\wedge\phi=0$.\qed

We will now present quite standard results on the invertible case. The reader may compare them to  Chow \cite{Cho}, in particular Theorem I and its proof, and to the results in \cite{Vil}. 

{\bf 8.5. Proposition.} Let $V$ and $W$ be two real vector spaces with same finite dimension $\geq 4$. If an invertible linear map $\lka:\bigwedge^2 V\to \bigwedge^2 W$ preserves decomposability then either there exist an invertible linear map $B: V \to W$ and a sign $\epsilon=\pm1$ such that for any $(q,v)\in V^2$, $\lka(q\wedge v)=\epsilon B(q)\wedge B(v)$, or $\dim V=\dim W=4$ and there are a nonzero $\mu\in\bigwedge^4W$ and an invertible linear map $C:V\to W^*$ such that $\lka(q\wedge v)=\bigl(C(q)\wedge C(v)\bigr)\lint\mu$.

{\bf 8.6.  Lemma.} An invertible ${\cal \lk}$ satisfying $(8.1)$ maps bijectively the decomposable bivectors of $V$ on the decomposable bivectors of $W$.

{\sl Proof.} As ${\cal \lk}$ is surjective, ${\cal \lk}^{(2)}$ reaches all the vectors of a standard basis of $\bigwedge^4
W$, i.e.\ ${\cal \lk}^{(2)}$ is surjective, and thus invertible. Clearly
$\bigl({\cal \lk}^{(2)}\bigr)^{-1}=({\cal \lk}^{-1})^{(2)}$, i.e.\ ${\cal \lk}^{-1}$ satisfies $(8.3)$ and thus $(8.1)$.\qed

{\bf 8.7.  Lemma.} Let $V$ be a vector space, $\dim V=n+1$. Let $H\subset\bigwedge^2 V$ be a vector subspace such
that (i) any
$\pi\in H$ is decomposable, (ii) if
a subspace $K\subset
\bigwedge^2 V$ contains $H$ and has the same property, then $H=K$. Either there exists $F\subset V$, $\dim F=3$,
such that
$H=\bigwedge^2 F$, or there exists a nonzero $v\in V$ such that $H=[v]\wedge V$. In the first case, $\dim H=3$. In the
second case, $\dim H=n$.

{\sl Proof.} Let $\xi_1$, $\xi_2,\dots$ be a base of $H$. The two-dimensional supports of $\xi_1$ and $\xi_2$ have a one-dimensional intersection $[x]\subset V$ (if it was zero-dimensional $\xi_1+\xi_2$ would not be decomposable, if it was
two-dimensional, $\xi_1$ and $\xi_2$ would be proportional). We write $\xi_1=x\wedge
y_1$ and $\xi_2=x\wedge y_2$. The support of $\xi_3$ must intersect both supports. Suppose first that the intersection is not $[x]$. Then $[\xi_1,\xi_2,\xi_3]=\bigwedge^2 F$ where $F=[x,y_1,y_2]$. There cannot be another bivector $\xi_4$ in the base of $H$, as the support of
$\xi_4$ would cut $F$ along a line, and a bivector $\eta\in[\xi_1,\xi_2,\xi_3]$ whose support does not contain this line is such that $\xi_4+\eta$ is not decomposable. Now if the intersection is
$[x]$, $\xi_3\in[x]\wedge V$. For $i>3$, $\xi_i\in[x]\wedge V$ by the same arguments.\qed

{\sl Proof of Proposition 8.5.} If  $n=\dim V-1\geq 4$, ${\cal \lk}$ defines a bijection ${\cal \lk}_{\cal P}:{\cal P}(V)\to{\cal P}(W)$ as follows. A nonzero vector $v\in V$ defines a line $[v]\subset V$ and an $n$-dimensional subspace $[v]\wedge V\subset \bigwedge^2V$. By Lemma 8.7, the image by ${\cal \lk}$ of such a subspace is a $[w]\wedge W$ for some nonzero $w\in W$. This induces the map ${\cal \lk}_{\cal P}: [v]\mapsto [w]$. This map is bijective, the inverse being constructed from ${\cal \lk}^{-1}$, which satisfies condition $(8.1)$ according to Lemma 8.6.

The bijection ${\cal \lk}_{\cal P}$ sends projective lines to
projective lines. To see this, observe that, $u_1$, $u_2$, $v$ being nonzero vectors in $V$, $u_1\wedge u_2\neq 0$,
the ``point" $[v]\in{\cal P}(V)$ belongs to the ``line'' $[u_1,u_2]$ if and only if $u_1\wedge u_2\in [v]\wedge V$. By
construction of ${\cal \lk}_{\cal P}$, the line $[u_1,u_2]$ is thus sent to the line corresponding to the decomposable bivector
${\cal \lk}(u_1\wedge u_2)$.

By the fundamental theorem of projective geometry (see e.g.\ \cite{Cho}, p.\ 33), as $\dim V=\dim W\geq 3$, there exists an invertible
linear map
$\tilde {\cal \lk}_{\cal P}: V\to W$ which maps ${\cal P}(V)$ on ${\cal P}(W)$ as does ${\cal \lk}_{\cal P}$ (we recall that our ground field is $\R$). The
relation between
${\cal \lk}$ and $\tilde {\cal \lk}_{\cal P}$ is
$$\hbox{for any } (u,v)\in V^2,\quad {\cal \lk}(u\wedge v)\wedge \tilde {\cal \lk}_{\cal P}(u)=0.\eqno(8.3)$$
For any $(u,v)\in V^2$, ${\cal \lk}(u\wedge v)$ and $\tilde {\cal \lk}_{\cal P}(u)\wedge \tilde {\cal \lk}_{\cal P}(v)$ are proportional bivectors. Each expression defines a linear map from $\bigwedge^2 V$ to $\bigwedge^2 W$. The proportionality factor between both maps must
be a constant. With the notation of $(8.2)$, ${\cal \lk}=\lambda (\tilde {\cal \lk}_{\cal P})^{\wedge 2}$, for some nonzero $\lambda\in\R$.

 If  $\dim V=4$, both types of maximal spaces described in Lemma 8.7 have dimension 3.
Each type form a connected sub-variety of the Grassmannian of $3$-planes in $\bigwedge^2 V$. Either
${\cal \lk}$ sends the $[u]\wedge V$'s to the $[v]\wedge W$'s as in higher dimension, or it sends the $[u]\wedge V$'s
to the $\bigwedge^2 F$'s (with $F\subset W$, $\dim F=3$.) In the first case, the conclusion is the same as in the case
$n\geq 4$. In the second case, we compose ${\cal \lk}$ with the bijection $\star:\bigwedge^2 W\to\bigwedge^2 W^*$, $\pi\mapsto
\pi\lint\mu$, where $\mu\in\bigwedge^4W^*$ is nonzero. The composed map $\star\circ {\cal \lk}:\bigwedge^2 V\to\bigwedge^2
W^*$ is in the first case. We have again the same conclusion for this map. This is the second case of Proposition 8.5.\qed

The second case of Proposition 8.5 is irrelevant in our study, and there are several ways to exclude it. We present here a proposition which looks like Proposition 8.5, but starts with a stronger hypothesis.

{\bf 8.8. Proposition.} Let $V$ and $W$ be two real vector spaces with same finite dimension $\geq 4$. If an invertible linear map $\lka:\bigwedge^2 V\to \bigwedge^2 W$ is such that for any $(q,v,w,x)\in V^4$, there exists a $z\in W$ such that $z\wedge \lka(q\wedge v)=z\wedge \lka(q\wedge w)=z\wedge \lka(q\wedge x)=0$, then  there exist an invertible linear map $B: V \to W$ and a sign $\epsilon=\pm1$ such that for any $(q,v)\in V^2$, $\lka(q\wedge v)=\epsilon B(q)\wedge B(v)$.

To compare the new hypothesis with $(8.1)$, we notice that $(8.1)$ may be expressed as: for any $(q,v,w)\in V^3$, $\lka(q\wedge v)\wedge\lka(q\wedge w)=0$. This is also: for any $(q,v,w)\in V^3$, there exists a $z\in W$ such that $z\wedge \lka(q\wedge v)=z\wedge \lka(q\wedge w)=0$.
In the new hypothesis the first two conditions uniquely determine $z$ as a function of $(q,v,w)$, and the last one shows that $\lka$ sends $[q]\wedge V\subset\bigwedge^2V$ on $[z]\wedge W\subset \bigwedge^2 W$. This excludes the second case in Proposition 8.5, as shown by the last argument in the proof of this proposition.

Equation $(7.7)$ shows that an invertible $\lk_{\cal B}$ satisfies the hypothesis of Proposition 8.8, with the restriction that $q$ should belong to the screen. But the hypothesis of Proposition 8.8 may be expressed as an algebraic identity, and thus may be extended to any $q\in V$.

\bigskip
\centerline{\bf 9. Extended Beltrami. The multidimensional case.}
\bigskip

There are several nearly equivalent ways of stating Beltrami's theorem. The statement in \cite{Mat} is quite elegant (compare \cite{Bel}, p.\ 204): if two Riemannian scalar products on a manifold are geodesically equivalent, and if one of them is of constant curvature, so is the other.

A scalar product of constant curvature is (locally) geodesically equivalent to a flat scalar product. Consequently, an alternative statement of Beltrami's theorem may start from an open domain of an affine space, and discuss the pseudo-Riemannian scalar products on it which have rectilinear geodesics. Here, since we extend the statement to the degenerate scalar products, we cannot speak of the geodesics of the scalar product. This is why our statements explicitly introduce a connection, and thus differ in appearance from the above elegant statement. 

Another unusual aspect of our statements is their conclusion, which describes a uniformizing screen, i.e.\ the target space for the central projection introduced in Lemma 7.2 and \S 7.3.  The non-metric construction of this screen fits with the degenerate cases. Furthermore, a uniformizing screen centrally projects on various affine spaces. The description of the uniformizing screen is shorter than the description of its projections.

{\bf 9.1. Proposition.} Consider a connected open set ${\cal U}$ of an $n$-dimensional affine hyperplane $A\subset V$, endowed with a connection which is geodesically equivalent to the affine connection. If a volume form and a scalar product of rank $\geq 2$ are invariant by parallel transport, then the uniformizing screen is part of an $n$-dimensional quadric centered at the center of projection.

{\bf Remark.} The rank one case shall be excluded, as is clear from the case $n=1$, where the uniformizing screen may be any curve transverse to the rays from the origin (the center of projection). Such a non-quadratic example generates another one in each dimension. For example, the cylinder that we got in Theorem 7.6 does not need to be quadratic. And, as soon as $n\geq 2$, a non-quadratic cylinder cannot be equivalent to a quadratic cylinder (see the last remark in \S 7.3).

Let us give a general result that implies Proposition 9.1 and includes the rank one case. We state it in two parts, the first part being a rather standard version of Beltrami's theorem.

{\bf 9.2. Theorem.} Consider a connected open set ${\cal U}$ of an $n$-dimensional affine hyperplane $A\subset V$, endowed with a connection which is geodesically equivalent to the affine connection. If a nondegenerate scalar product is invariant by parallel transport, then the uniformizing screen ${\cal H}$ is either 

(i) part of a nondegenerate centered quadric in $V$: there exist a nondegenerate quadratic form $G$ on $V$ and a nonzero $\lambda\in\R$ such that ${\cal H}$ is included in the hypersurface with equation $G(q)=1$ and the scalar product on ${\cal H}$ is defined by the quadratic form $v\mapsto \lambda G(v)$ for any $q\in {\cal H}$ and $v\in
T_q{\cal H}$, or

(ii) part of an affine hyperplane: there exist a nonzero $\phi\in V^*$ and a nondegenerate quadratic form $g$ defined on $\ker\phi$ such that ${\cal H}$ is included in the hyperplane with equation $\la
\phi,q\ra=1$ and the scalar product on ${\cal H}$ is defined by $g$, or

(iii) in the case $n=1$, part of an arbitrary curve transverse to the rays in $V$.

{\bf 9.3. Theorem.} Consider a connected open set ${\cal U}$ of an $n$-dimensional affine hyperplane $A\subset V$, endowed with a connection which is geodesically equivalent to the affine connection. If a volume form and a scalar product of rank $k$, $k\ge 1$, are invariant by parallel transport, then the uniformizing screen is part of an $n$-dimensional cylinder whose generating affine subspaces have dimension $n-k$. On the screen, the degeneracy spaces of the scalar product are these generating subspaces. Consider the canonical projection $V\to V/N$, where $N$ is the $n-k$ dimensional vector space, direction of the generating subspaces. The cylinder and its scalar product project on a model described in Theorem 9.2.

{\bf 9.4. The proofs.} We use the notation and repeat the arguments of \S 7.7. We deduce the compatibility condition $(7.7)$ between the quadrilinear form $\lk_{\cal A}$ encoding the scalar product and the function $h$ such that $h(q)=1$ is the equation of ${\cal H}$.

We first prove Theorem 9.3. We consider any element $k$ of the absolute kernel $N$, i.e.\ such that $\lk_{\cal A}(k,.;.,.)=0$, and follow the deduction at the end of \S7.8. For all $q\in {\cal H}$, $\la dh|_q,k\ra=0$. The screen ${\cal H}$ is locally translation invariant in the direction $N$. Being the absolute kernel, this direction is a degeneracy for the scalar product at any $q\in{\cal H}$.

Consider now the quotient space $V/N$. The quadrilinear form $\lk_{\cal A}$ is the pull-back of a quadrilinear form on $V/N$, with trivial absolute kernel. The uniformizing screen ${\cal H}$ is part of the pull-back of a screen in $V/N$. The compatibility condition $(7.7)$ passes to the quotient. On the quotient space $V/N$ we have the conditions of Theorem 9.2, except that the hypothesis of nondegeneracy of the scalar product is changed into the triviality of the absolute kernel $N$.

To conclude the proof of Theorem 9.3, we should consequently get the conclusions of Theorem 9.2 by using the assumption $N=\{0\}$ instead of the nondegeneracy of the scalar product. As we just showed that a nontrivial $N$ is a degeneracy, Theorem 9.2 will be proved at the same time. We begin with a lemma, which shows that in most cases there is at most one uniformizing screen compatible with a given $\lk_{\cal A}$.

{\bf 9.5. Lemma.} Consider two non-homothetic screens in the $n+1$-dimensional real vector space $V$, cutting a common open set of rays. If the same quadrilinear form $\lk_{\cal A}: V^4\to \R$ with a $2\times 2$ Young tableau symmetry induces on each screen a scalar product which is parallel for the central connection, then there exist a $\lambda\in\R$ and a $(\xi,\eta)\in (V^*)^2$ such that $\lk_{\cal A}=\lambda\, \xi\wedge \eta\otimes\xi\wedge\eta$.

{\sl Proof.} Let $h(q)=1$ and $k(q)=1$ be the equations of the two screens, where $h$ and $k$ are positively homogeneous functions (for example of degree 1). According to $(7.7)$, for any $v\in V$,  for any $q$ such that the ray $[q]$ cuts both screens, we have $dh|_q\wedge \lk_{\cal A}(q,v;.,.)=0$ and $dk|_q\wedge \lk_{\cal A}(q,v;.,.)=0$. If furthermore $q$ is such that $dh|_q\wedge dk|_q\neq 0$, then there is a $\lambda$ depending on $(q,v)$ such that $\lk_{\cal A}(q,v;.,.)=\lambda dh|_q\wedge dk|_q$. The left-hand side is skew symmetric in $(q,v)$. Exchanging $q$ and $v$ we see that the right-hand side does not depend on $q$, except for a factor. There are an $\omega\in \bigwedge^2V^*$ and a $(\xi,\eta)\in(V^*)^2$ such that $\lk_{\cal A}(q,v;.,.)=\la \omega,q\otimes v\ra \xi\wedge\eta$, on the relevant domains and consequently everywhere. Thus $\lk_{\cal A}=\omega\otimes \xi\wedge\eta$. We get the result from the symmetry of $\lk_{\cal A}$ seen as a bilinear form on bivectors.\qed

As we just explained, we assume $N=\{0\}$. In the case $\lk_{\cal A}=\lambda\, \xi\wedge \eta\otimes\xi\wedge\eta$, $N$ is the intersection of $\ker \xi$ and $\ker \eta$, which is nontrivial as soon as $\dim V=n+1\geq 3$. The screen is unique (up to homothety) except if $n=1$. The case $n=1$ is the obvious case (iii) of Theorem 9.2.

As the case $n=2$ is already treated by Theorem 7.6, we assume $n\geq 3$ and consider the map $\lk_{\cal B}:u\wedge v\mapsto \lk_{\cal A}(u,v;.,.)$, which preserves decomposability according to $(7.7)$.

As a first case, we assume that this map is non-invertible. Proposition 8.1 shows the existence of a nonzero $\phi\in V^*$ such that $\phi\wedge \lk_{\cal A}(u,v;.,.)=0$ for any $(u,v)\in V^2$. We compare this equation with $(7.7)$ and call ${\cal H}$ the affine hyperplane with equation $\la \phi, q\ra=1$. The compatibility condition $(7.7)$ is satisfied, ${\cal H}$ is the uniformizing screen (unique according to Lemma 9.5), the central connection is the affine connection and $\la u,v\ra = \lk_{\cal A}(q,u;q,v)$ is the translation invariant scalar product, which is nondegenerate due to $N=\{0\}$. This is case (ii) of Theorem 9.2. 

The other case is of an invertible $\lk_{\cal B}$. By Proposition 8.8 and the argument that follows it, there is a linear map $B:V\to V^*$  such that $\lk_{\cal B}=\pm B^{\wedge 2}$. Theorem {7}.{5} in \cite{Vil} shows that $B$ is symmetric, as a consequence of the algebraic Bianchi identity satisfied by $\lk_{\cal A}$. Equation $(7.7)$ becomes $dh|_q\wedge B(q)\wedge B(v)=0$. This equation is satisfied if we set $h(q)=\la B(q),q\ra$. Thus a level hypersurface of $h$ is a uniformizing screen such that the scalar product $\la u,v\ra=\lk_{\cal A}(q,u;q,v)$ is parallel for the central connection. Such screen is unique up to homothety according to Lemma 9.5. We have $\la u,v\ra=\la \lk_{\cal B}(q\wedge u),q\wedge v\ra=\pm\la B(q)\wedge B(u),q\wedge v)=\pm\la B(q),q\ra\la B(u),v\ra\mp\la B(q),v\ra\la B(u),q\ra$. As $u$ and $v$ are tangent vectors, $\la B(q),v\ra=\la B(u),q\ra=0$. Thus $\la u,v\ra$ is, up to a constant factor, $\la B(u),v\ra$. This is case (i) of Theorem 9.2, where $G(v)=\pm\la B(v),v\ra$.\qed

\bigskip
\centerline{\bf 10. The degenerate case. Examples and properties.}
\bigskip

If, starting with a Newton system $\ddot q=f(q)$, we find a triple $(\mu, g, U)$, consisting of a function $\mu(q)$ defining a change of time, a {\it nondegenerate} scalar product $g$, and a potential function $U(q)$, such that $Dg=0$ and $\mu^{-2}f=\grad_g U$, then we have a quadratic first integral, the energy, and a conserved symplectic form. This is what we explained in the introduction.

In Sections 2 to 6, we learned how to deal with the leading term of a polynomial first integral\footnote{Laplace \cite{Lap} gave what is maybe the first method of determination of the first integrals of a given system. Considering the Kepler problem in space, he devised a systematic research of polynomial first integrals, and developed it until he found all the classical first integrals. These are at most quadratic in the
velocities. Among them is the so-called Laplace-Runge-Lenz or eccentricity vector. One could believe that Laplace discovered it in this way. Actually he had read 
a famous work by Lagrange \cite{Lag} where the formulas for the coordinates of this vector appear. Nevertheless, Laplace's method is interesting and general. The first step is to get an {\it a priori} information on the highest degree term of the first integral. After many authors, we pushed this part of his study forward in our sections 2 to 6 (compare for example \cite{Whi}, p.\ 332).}. In Section 7 to 9, we expressed within this framework the special properties of a quadratic first integral that may be, after a change of time, an energy first integral. We introduced the uniformizing screen associated to such a quadratic first integral.

Our studies include the case of a degenerate scalar product $g$. We have shown how simple is the classification of such objects. The bi-quadratic polynomial defining $g$ has an absolute kernel and the uniformizing screen has a cylindric direction.

The degenerate case includes some interesting examples. In the generalized planar Kepler problem, the square of the angular momentum produces the cylindric screens which are introduced in \cite{Al1}. This is the case of Lemma 9.5 where the uniformizing screen is not unique.

Another interesting example is a potential defined on the affine space $A$ by a formula $q\mapsto U(q-a)$, where $a\in A$ and where $U$ is a function which is positively homogeneous of degree $-2$. This includes the $n$-body problem with $1/r^3$ law of force, which was studied within this general framework by Jacobi in 1842 (see some references in \cite{Al2}). There is the quadratic first integral $G(q,\dot q)=2\|q-a\|^2H-\la q-a,\dot q\ra^2$, where $H(q,\dot q)=\|\dot q\|^2/2-U(q-a)$ is the energy. After homogeneization by Formula $(2.2)$, the leading term of the first integral is $\|u\wedge \dot u\|^2$, where $u=q-q_0a$. The vector line $[a]$ is the absolute kernel of this expression. It is also the direction of the generating lines of a cylindric uniformizing screen in $V$, which cuts the affine hyperplane $A$ along the sphere $\|q-a\|=1$. This example is interesting in itself\footnote{A global aspect of projective dynamics is the possible extension ``after infinity" of the orbits in a natural system. The first example is the equation $\ddot
q=0$, where a rectilinear trajectory is extended into the double covering of a projective line. In the two-body problem the extension is as
expected: any hyperbolic trajectory is extended into an ellipse. We showed in \cite{Al1} that the
motion of a particle in the gravitational field of any fixed distribution of mass may also be extended after infinity. In the three-body problem, the possibility of an extension after infinity was discussed by
Chazy (see \cite{Cha},  p.\ 52, and second footnote, p.\ 59), who obtained a rather negative answer: only some particular hyperbolic trajectories can be continued. In our language the
projective force field becomes singular at infinity. Chazy noticed that if we define the interaction between the bodies by an inverse cube law instead
of the Newtonian inverse square law, the singularities disappear. There is an extension of all the unbounded orbits. This is also what we see directly on the cylindric screen associated to the first integral $G$. The orbits cross the hyperplane ``at infinity'', which is materialized as the vector hyperplane parallel to the affine hyperplane $A$.}, but also as a case which is not integrable in general, although there are two independent quadratic first integrals $H$ and $G$.

We will conclude our study by showing that the conserved symplectic form of the nondegenerate case has an analogue in the degenerate case, which is a conserved pre-symplectic form (i.e.\ a conserved closed 2-form).

{\bf 10.1.} Let $M$ be an $n$-dimensional manifold, thought of as the configuration space of a mechanical system. 
From a local chart
$$\Phi:\Omega\longrightarrow\R^n,\qquad q\longmapsto \Phi(q)=(x_1,\dots,x_n),$$
of $M$, where $\Omega\subset M$ is an open set, we form an {\it adapted local chart} of the tangent bundle $TM$:
$$\Phi_*:T\Omega\longrightarrow\R^{2n},\qquad \xi\longmapsto
\bigl(\Phi(q),d\Phi|_q(\xi)\bigr)=(x_1,\dots,x_n,y_1,\dots ,y_n),$$ where $T\Omega\subset TM$ is the inverse image of
$\Omega$ by the canonical projection $TM\to M$, and  $q\in \Omega$ is the image of $\xi\in T\Omega$ by this projection.

{\bf 10.2.} A {\it second order differential equation} on a manifold $M$ is a vector field $Z$ on $TM$ such that for any adapted chart $(x_1,\dots,x_n,y_1,\dots ,y_n)$ and for any $i$, $1\leq i\leq n$, we have $\partial_Z x_i=y_i$. By $\partial_Z x_i$ we mean the derivative of the
function $x_i$ along the vector field $Z$.

{\bf 10.3.} A {\it pre-Lagrangian} for the second order differential equation $Z$ is a function $L: TM\to \R$ such that
in any adapted chart $(x_1,\dots,x_n,y_1,\dots,y_n)$ the Lagrange equations $$\partial_Z\Bigl(\frac{\partial L}{\partial
y_i}\Bigr)-\frac{\partial L}{\partial x_i}=0$$
are satisfied. Here the derivation symbol $\partial_Z$ may be replaced by the more familiar notation $d/dt$.

We do not require that the Lagrange equations define the vector field $Z$. Our only requirement is that these equations are true. The pre-Lagrangians form a vector space, which includes the functions $\sum_{i=1}^n\eta_i
y_i$, where the $\eta_i$'s are the coordinates of a closed $1$-form on $M$. An example of second order differential
equation with many quadratic pre-Lagrangians
 is the harmonic oscillator on
$\R^n$, with equations
$\ddot x_i=-x_i$,
$1\leq i\leq n$. The functions
$y_iy_j-x_ix_j$, $1\leq i\leq j\leq n$, are pre-Lagrangians.

The following proposition
is well-known in the case of a Lagrangian, and the proof is not more complicated in the case of a pre-Lagrangian.

{\bf 10.4.  Proposition.} Let $L$ be a pre-Lagrangian for a second order differential equation $Z$. Given an adapted
chart
$(x_1,\dots,x_n,y_1,\dots,y_n)$, let
$p_i=\partial L/\partial y_i$. The function $\sum_{i=1}^n p_iy_i$ does not depend on the chart and
$E=\sum_1^n p_iy_i-L$ is a first integral of $Z$.

{\bf 10.5.  Proposition.} Let $Z$ be a second order differential equation and $G: TM\to  \R$ be a function. We choose
a local chart
$(x_1,\dots,x_n)$ of $M$ and define locally
$G_*: TM\to T^*M,\; (x_1,\dots,x_n,y_1,\dots,y_n)\mapsto (x_1,\dots,x_n,p_1,\dots,p_n)$, where
$p_i=\partial G/\partial y_i$.
Let $\omega$ be the pre-symplectic form on $TM$, pull-back by $G_*$ of the canonical
symplectic form on $T^*M$. Let $\Omega\subset M$ be a simply connected open set and $T\Omega\subset TM$ its inverse
image by the canonical projection $\pi:TM\to M$. The Lie derivative
${\cal L}_Z\omega$ vanishes on $T\Omega$ if and only if there exists a $U:\Omega\to \R$ such that
$L=G+U\circ\pi$ is a pre-Lagrangian of
$Z$ on
$T\Omega$.

{\sl Proof.} We write
$$p_i=\frac{\partial G}{\partial y_i},\quad\omega=\sum_id p_i\wedge dx_i,\quad Z\lint\omega=
\sum_i(\partial_Zp_i)dx_i-y_i dp_i,$$
$$\sum_i y_idp_i=d\Bigl(\sum_iy_ip_i\Bigr)-\sum_ip_idy_i=d\Bigl(\sum_iy_ip_i-G\Bigr)+\sum_i\frac{\partial
G}{\partial x_i}dx_i.$$
Erasing the exact form, and using the Cartan formula ${\cal L}_Z\omega=d(Z\lint\omega)$ we conclude that $\omega$ is preserved if and only if
$$\sum_i\bigl(\partial_Zp_i- \frac{\partial
G}{\partial x_i}\bigr)dx_i$$
is closed. In particular, the coefficients must be independent of the $y_i$'s. Locally there exists a function $U(q)$
such that
$$\partial_Z p_i=\frac{\partial (G+U)}{\partial x_i},$$
which are the Lagrange equations for the pre-Lagrangian $G+U$.\qed

{\bf Remark.} If $M$ is not simply connected and ${\cal L}_Z\omega=0$ on $TM$, then there exists a multivalued
pre-Lagrangian for $Z$, of the form $G+U\circ \pi$, where  $U$ is a
multivalued function on $M$.

{\bf 10.6. More specific systems.} Recall that we denote by $\xi\in TM$ the state of the particle and $q\in M$ its
position, i.e.\ the canonical projection of $\xi$. Among the second order equations on $M$, those of the form
$$D_\xi\xi=f(q),\eqno(10.1)$$
where $D$ is a symmetric (also called torsion-free) linear connection, and $f$ is a tangent vector field on $M$, have simple properties. Here
$\xi$ also denotes the velocity field along the trajectory. Let $L$ be a pre-Lagrangian for this equation, which is a
polynomial in the velocity. Then both the even part and the odd part of $L$ are polynomial pre-Lagrangians. If $L$ is even
and of degree two, then
$L=T+U$, where
$T$ is a quadratic form in the velocity $\xi$, and $U$ only depends on the position $q$. There exists a scalar product $g$ on $M$ such that $2T(\xi)=g(\xi,\xi)$. The Lagrange equations are equivalent to: $Dg=0$ and $g(f,.)=dU$.

{\bf 10.7. Screen dynamics.} A system constrained to a screen ${\cal H}$, of the general form
$${d^2 q\over dt^2}=f(q)+\lambda q,\eqno(10.2)$$ as for example $(7.2)$, is of
type $(10.1)$. A system $(10.2)$ where $f(q)$ is not tangent to the screen ${\cal H}$ is obviously equivalent to another one with $f(q)$ tangent to the screen, due to the multiplier $\lambda$ in the right-hand side. When comparing $(10.2)$ with
$(10.1)$, we assume that $f(q)$ is tangent to ${\cal H}$.

A symmetric connection $D$ on ${\cal H}$ is induced by the standard affine connection on $V$ and by the splitting $T_qV=T_q{\cal H}\oplus [q]$ at any $q\in{\cal H}$. In Definition 7.1, we called this connection  the central connection on the screen ${\cal H}$. System $(10.2)$ is $D_\xi\xi=f(q)$, where $\xi$ is the velocity field along the trajectory. The above conclusions apply, as well as the conclusions of Theorem 9.3.

{\bf 10.8. Theorem.} If System $(10.2)$, defined by a screen ${\cal H}$ and a force field $f$ tangent to ${\cal H}$,  possesses a quadratic pre-Lagrangian $L=T+U$,
the free motion $\ddot q=\lambda q$ on ${\cal H}$ possesses the quadratic pre-Lagrangian $T$.  Call $g$ the (possibly degenerated) scalar product on ${\cal H}$ such that $2T=g(\dot q,\dot q)$, and $D$ the central connection on ${\cal H}$. The Lagrange equations are equivalent to: $Dg=0$ and $g(f,.)=dU$. The energy $T-U$ is a first integral, and the pre-symplectic form $\omega$ of Proposition 10.5 is conserved by the flow. If $g$ is nontrivial, the screen ${\cal H}$ is part of a cylinder. The degeneracy spaces of $g$ are the generating subspaces of the cylinder. Consider the canonical projection $V\to V/N$, where $N$ is the direction of the generating subspaces. The cylinder and its scalar product project on one of the models of constant curvature listed in Theorem 9.2.

{\bf Acknowledgements.} I wish to thank Alain Chenciner for the idea of using $(2.5)$ to reprove Nijenhuis theorem (Proposition 2.3), Laurent Niederman for indicating me works on separately polynomial functions, Thierry Combot for convincing me not to try to generalize Proposition 2.4 as suggested by Proposition 2.3, and for other useful information. I wish to thank Bartolom\'e Coll for, among other crucial information, indicating me the works by Nijenhuis and Thompson, Jos\'e Mar\'\i a Pozo Soler for indicating me the footnote in Penrose-Rindler's book, for pointing out an important missing case in my first proof of Theorem 3.4, and for giving me valuable advices. I wish to thank Hans Lundmark for his precious help about all the aspects of this work, and Konrad Sch\"obel for showing me the work of McLenaghan, Milson and Smirnov. I wish to thank Vladimir Matveev and Abdelghani Zeghib for their indications about Beltrami's theorem and related topics. I wish to thank H.\ Scott Dumas, Mauricio Garay, Richard Montgomery, Maria Przybylska, and Pierre Teyssandier for their continuous help and stimulating comments. Finally, I wish to thank Alexey Borisov and Ivan Mamaev for many discussions and accurate reactions concerning this topic in the past ten years.

\end{document}